\documentclass[
 reprint,
superscriptaddress,
showpacs,preprintnumbers,
 amsmath,amssymb,
aps,
pra,
]{revtex4-1}
\usepackage[dvipsnames]{xcolor}
\usepackage[colorlinks=true,bookmarks=false,linkcolor=NavyBlue,urlcolor=NavyBlue,citecolor=NavyBlue,breaklinks]{hyperref}
\usepackage{graphicx}
\usepackage{dcolumn}
\usepackage{bm}

\begin{document}

\title{Zitterbewegung-mediated RKKY coupling in topological insulator thin films}

\author{Cong Son Ho}
 \email{sonhc85@gmail.com}
 \affiliation{
Department of Electrical and Computer Engineering, National University of Singapore,4 Engineering Drive 3, Singapore 117576, Singapore.
}

 \affiliation{
	Chemical and Quantum Physics, School of Science, RMIT University, Melbourne, Australia
}

\author{Seng Ghee Tan}
\affiliation{
Department of Optoelectric Physics, Chinese Culture University, 55 Hwa-Kang Road, Yang-Ming-Shan, Taipei 11114, Taiwan
}

\author{Zhuo Bin Siu}

 \affiliation{
Department of Electrical and Computer Engineering, National University of Singapore,4 Engineering Drive 3, Singapore 117576, Singapore.
}

\author{Mansoor B. A. Jalil}%
\email{elembaj@nus.edu.sg}
\affiliation{
Department of Electrical and Computer Engineering, National University of Singapore,4 Engineering Drive 3, Singapore 117576, Singapore.
}


\begin{abstract}
The dynamics of itinerant electrons in topological insulator (TI) thin films is investigated using a multi-band decomposition approach. We show that the electron trajectory in the 2D film is anisotropic and confined within a characteristic region. Remarkably, the confinement and anisotropy of the electron trajectory are associated with the topological phase transition of the TI system, which can be controlled by tuning the film thickness and/or applying an in-plane magnetic field. Moreover, persistent electron wavepacket oscillation can be achieved in the TI thin film system at the phase transition point, which may assist in the experimental detection of the jitter motion (Zitterbewegung). The implications of the microscopic picture of electron motion in explaining other transport-related effects, e.g., electron-mediated RKKY coupling in the TI thin film system, are also discussed.
\end{abstract}

\maketitle

\section{Introduction}

The Zitterbewegung (ZB) effect, first described by Schr\"odinger \cite{schrodinger1930kraftefreie}, refers to the oscillatory motion of relativistic free electrons which is governed by the Dirac equation. Recently, with the rapid advance of spintronics, the ZB effect has been also studied in various systems including spin-orbit coupling (SOC) systems \cite{Schli05,schliemann2006zitterbewegung,demikhovskii2008wave,biswas2012zitterbewegung,biswas2014wave, Shi13}, monolayer and bilayer graphene \cite{Rusin07,PhysRevB.78.125419,PhysRevB.78.235321,Rusin09,Wang10,martinez2010klein, Shi13}, carbon nanotubes \cite{Rusin14}, topological insulators \cite{Shi13,yanyushkina2012zitterbewegung,PhysRevB.98.165120}, Weyl semimetals \cite{huang2018zitterbewegung}, superconductors \cite{Cannata90,Lurie70}, and ultra-cold atoms \cite{leblanc2013direct, Zhang13, PhysRevA.88.021604, Vaish08}. Experimentally, the ZB effect has only been directly observed in trapped ion \cite{Lamata07,Bermu07,Gerrit10,Qu13} and ultra-cold atomic systems \cite{leblanc2013direct, PhysRevA.88.021604}, and indirectly in solid state systems \cite{stepanov2016coherent,iwasaki2017observation} . The obstacles in observing this elusive phenomenon are due to the intrinsic properties of high oscillation frequency and rapid damping \cite{huang1952zitterbewegung, lock1979zitterbewegung, Schli05,Rusin07}.

In general, the ZB frequency scales with the energy gap and can be reduced in systems with narrow energy gaps, e.g., narrow gap semiconductors \cite{PhysRevB.72.085217} and topological insulators \cite{Shi13}. At the same time, the oscillation of a wavepacket usually decays over time, which results from the interference between oscillations of different momentum-dependent frequencies \cite{Schli05,Rusin07}. Therefore, for the ZB effect to be observed, it is crucial to prolong or even indefinitely sustain the oscillatory motion. There have been some proposals to achieve persistent ZB motion, for example, by using semiconductor nanowires \cite{Schli05}, or time-dependent systems \cite{ho2014persistent, PhysRevB.101.094306}. In principle, we can also design a system in which the ZB oscillation frequency is independent of electron momentum. In this way, we can avoid the interference effect and render the ZB motion persistent and robust against damping. 

In this work, we show that such persistent ZB motion can be realized in topological insulator (TI) thin films \cite{zyuzin2011parallel,Linder:prb09,Lu:prb10,Liu:prb10}. TI thin films differ from the more commonly studied semi-infinite TI slabs in that they have both a top and bottom surface, each of which can host surface states. The surface states on the two surfaces are coupled to each other due to the finite thickness of the film. In such thin films, the energy gap in the surface states can be controlled by applying an in-plane magnetic field \cite{zyuzin2011parallel} or tuning the thickness of the film \cite{Linder:prb09,Lu:prb10,Liu:prb10}. Topological phase transitions can thus be induced by closing the gap. We show that at the transition point,  there exists a momentum-independent oscillation frequency, which can give rise to persistent ZB oscillations of electron wavepackets. Furthermore, we find that the motion of electron in the $x-y$ plane is anisotropic with respect to the injection direction and confined to a certain region of the TI film.The anisotropy of the electron motion due to the ZB effect has consequences for transport-related properties of the thin film system.

 Here, we focus on the inter-layer interaction between two localized magnetic centers by means of Ruderman-Kittel-Kasuya-Yosida (RKKY) mechanism \cite{RKKY1,RKKY2,RKKY3,mattis2006theory}.  The RKKY interaction has been extensively investigated  in various systems such  as  superconductors \cite{TAGIROV1993257,akbari2011rkky,aristov1997rkky}, topological insulators \cite{RKKYTI:PRL2011,PhysRevB.81.233405,PhysRevB.81.172408,PhysRevLett.106.097201,PhysRevLett.106.136802,PhysRevB.90.125443}, Weyl and Dirac semimetals \cite{PhysRevB.92.241103,Sun_2017,PhysRevB.101.085419,PhysRevB.99.165302,PhysRevB.93.094433,PhysRevB.92.224435}, graphene  \cite{PhysRevB.76.184430,PhysRevLett.101.156802,PhysRevB.81.205416,PhysRevB.83.165425},  carbon  nanotubes  \cite{PhysRevLett.102.116403,PhysRevB.87.045422},   semiconductor quantum wires \cite{PhysRevB.81.113302,PhysRevB.79.205432}, and tunneling junctions \cite{PhysRevB.54.12953}. The RKKY interaction is mediated by the itinerant electrons. Intuitively, one would then expect the enhancement of the RKKY interaction when the magnetic centers lie along a preferred direction of electron motion, and a corresponding suppression of the RKKY interaction when the electrons are prohibited from moving between the two centers. We find that, indeed,  the anisotropy of the RKKY coupling is in line with that of the electron motion. We show that maximum RKKY coupling occurs when the separation between the two magnetic centers is perpendicular to the line connecting the Dirac points. 

This manuscript is organized as follows. In section \ref{sec2}, we present the model Hamiltonian and derive the dynamics of both plane-wave and wavepacket electrons. We discuss the confinement of the electron trajectory and the regime conditions for persistent ZB oscillation.  In section \ref{sec3}, the RKKY coupling is calculated in both the weak and strong hybridization limits, and its correlation with the electron motion is also discussed. Finally, section \ref{sec4} contains a summary of our main conclusions.

\section{Electron dynamics}\label{sec2}
\begin{figure}
\includegraphics[width=0.47\textwidth]{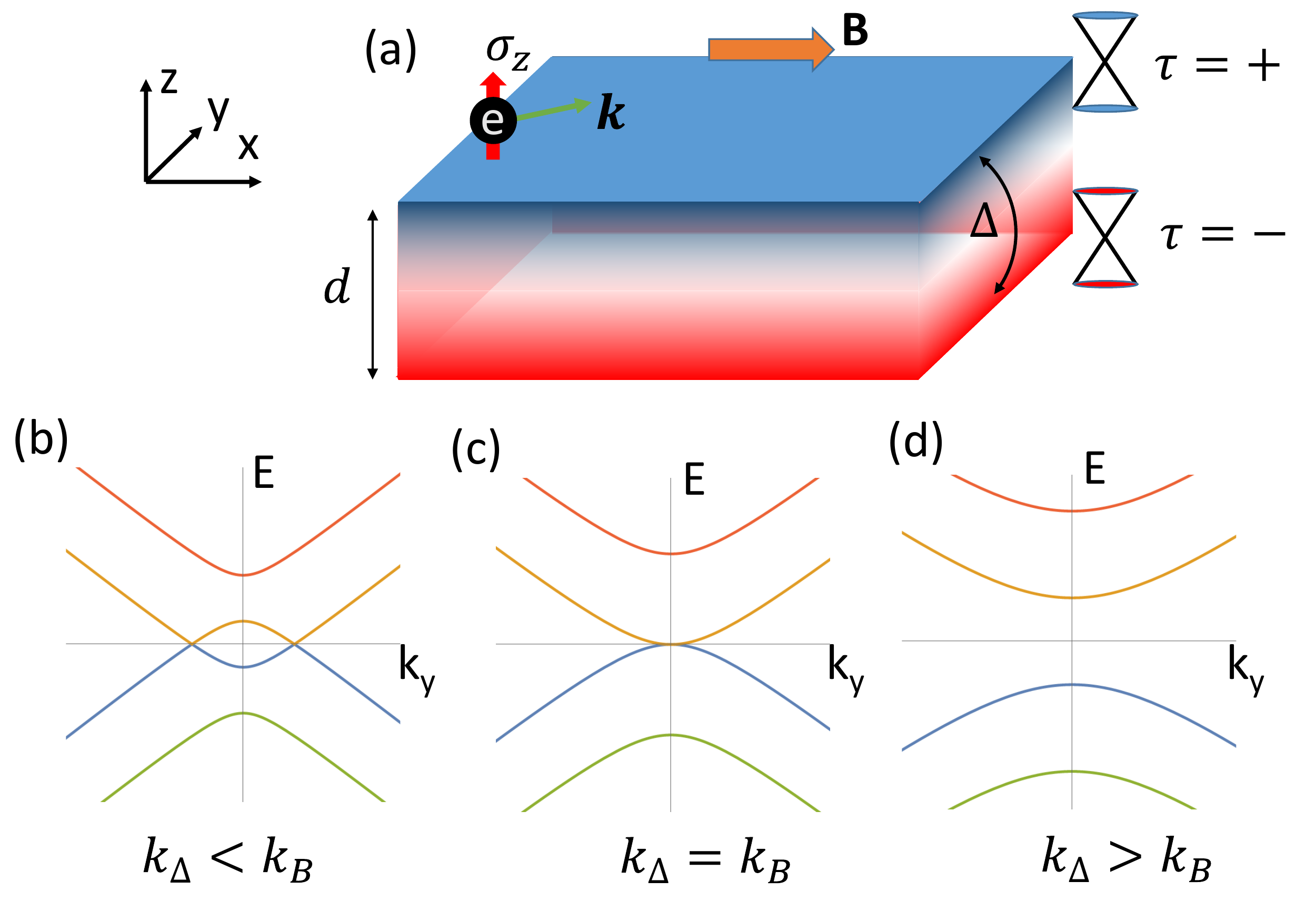}
\caption{(a) Schematic diagram of a thin TI film subjected to an in-plane magnetic field $\bm B$. The top and bottom surfaces are hybridized with hybridization energy $\Delta$. (b)--(d) Energies along $k_y$ of the TI thin film at different hybridization energy at $k_x=0$. The energy gap depends on the relative value of $k_\Delta$ and $k_B$, which are the characteristic momenta representing the hybridization and magnetic energies, respectively.}
\label{Fig1}
\end{figure}

We first consider a TI thin film subject to an in-plane magnetic field. For simplicity, we assume that the magnetic field is applied along the $x$-direction, so that the corresponding gauge field is ${\bf\mathcal{A}}_B=-\hat{{\bm y}}Bz$. As the thickness $d$ of the thin film is comparable to the surface state decay length, the two surfaces are hybridized. The effective Hamiltonian of the system is then \cite{zyuzin2011parallel}
\begin{equation}\label{Hal}
H_0= \tau_z H_D({\bm k}-\tau_z{\bm k}_B)+ \tau_x\Delta , 		
\end{equation}
where $H_D({\bm k})=\hbar v_f( {\bm z} \times {\bm \sigma} )\cdot{\bm k}$ is the Dirac Hamiltonian describing the topological surface state, in which $v_f$ is the Fermi velocity, ${\bm \sigma}$ the vector of Pauli spin matrices, and ${\bm z}$ the unit vector perpendicular to the film (see Fig. \ref{Fig1}). $\Delta$ is the hybridization parameter describing the coupling between the top and bottom surfaces, and $\bm \tau$ the vector of the Pauli matrices in pseudo-spin space that represents the electron occupancy at the top and bottom surfaces. For simplicity, we set $\hbar=1$, and introduce the characteristic momenta corresponding to the hybridization energy $k_\Delta=\Delta/v_f$ and the wavevector corresponding to the magnetic field ${\bm k}_B=e/2c Bd\hat{\bm{y}}$. The eigenenergies of the system are then given by
\begin{eqnarray}
E^s_{\tau}=sv_fk_\tau\label{eigen},\ \ k_\tau=\sqrt{k_u^2+k_v^2+ 2\tau k_uk_v\sin{\Phi}},
\end{eqnarray}
in which $s,\tau=\pm$ represent the real spin and pseudo-spin indexes respectively, and we define $k_u=\sqrt{k_\Delta^2+k_y^2}, k_v=\sqrt{k_B^2+k_x^2},  \Theta=\arctan{\frac{k_y}{k_\Delta}},$ and $\Phi=\arctan{\frac{k_B}{k_x}}$. The bandstructure of the TI film is depicted in Fig. (\ref{Fig1}). An energy gap of $E_g=v_f(k_\Delta-k_B)$ is formed when $k_\Delta>k_B$. Otherwise, the bandstructure is gapless, with the formation of two Dirac cones separated by $2q_0=2\sqrt{k_B^2-k_\Delta^2}$ along the direction perpendicular to the magnetic field. In particular, at the transition value $k_\Delta=k_B$, the two Dirac cones merge to form a single cone. 

The corresponding eigenstates are given by the four-vectors 
\begin{eqnarray}
|\psi_{s\tau}\rangle=N_{s\tau}
\left[\begin{array}{c}f_{s\tau}\\ g_{s\tau}
\end{array}\right]
\end{eqnarray}
where $N_{s\tau}$ are the normalization factors, and 
\begin{eqnarray}
f_{s\tau}&&=(\tan\Theta-\tau\sec\Theta)\left[\begin{array}{c}-s\frac{\tau k_u-ie^{i\Phi} k_v}{k_\tau}\\1\end{array}\right],\\  g_{s\tau}&&=\left[\begin{array}{c}1\\-s\frac{\tau k_u-ie^{i\Phi} k_v}{k_\tau}\end{array}\right].
\end{eqnarray} 

To study the ZB in this multi-band system, we derive the time-evolution of position operator, which is described in the Heisenberg picture as ${\hat{\bm r}}(t)=e^{iH_0t}{\hat{\bm r}}(0) e^{-iH_0t}$, and which at $t=0$ is formally represented by ${\hat{\bm r}}(0)=i\nabla_{\bm k}$. The time-dependent position operator comprises of a non-oscillatory part that describes the translational motion related to the intraband interference, and an oscillatory part that is associated with the ZB motion  \cite{Schli05,Rusin07,david2010general,ZB:review1} and related to the interband interference. Our interest lies in the latter, which is given by \cite{david2010general}
\begin{eqnarray}\label{ZB1}
{\hat{\bm r}}_Z ({\bm k},t)=\sum_{i\ne j} {\hat{\bm r}}_{ij}({\bm k},t), \ \ {\hat{\bm r}}_{ij}({\bm k},t)={\hat{\bm A}}_{ij}e^{i \Omega _{ij} t}, 	
\end{eqnarray}
in which $\Omega _{ij}=(E_i-E_j)$, with $i=(s,\tau)$  and ${\hat{\bm A}}_{ij}=i{\hat Q}_i \nabla_{\bm k} {\hat Q}_j$ are the frequency and amplitude of the oscillation, respectively.  In the above, we have introduced projection operators ${\hat Q}_{s\tau} =|\psi_{s\tau}\rangle\langle \psi_{s\tau}|$, so that the Hamiltonian (\ref{Hal}) can be decomposed as $H_0=\sum_{s\tau}E_{s\tau} {\hat Q}_{s\tau} $. We can further express the projection operators as
\begin{equation}
{\hat Q}_{s\tau}=\frac{1}{4} (1+s{\hat R})(1+\tau{\hat T}) \label{Qst},
\end{equation}
where $\hat{R}$ and $\hat{T}$ are involution operators satisfying ${\hat R}^2={\hat T}^2=1,[{\hat R},{\hat T}]=0$. The explicit forms of these operators are given in Appendix (\ref{App1}).
Due to the electron-hole symmetry of the eigenenergies given in Eq. \eqref{eigen}, there are only four distinct beat frequencies corresponding to the differences between the energies of interfering eigenstates. These frequencies are given by
\begin{eqnarray}\label{freq}
w_\pm=2v_fk_{\pm}, \ \ w_1=v_f(k_++ k_-),\ \ w_2=v_f(k_+- k_-),
\end{eqnarray}
where $k_\pm$ is given by Eq. \eqref{eigen}.

\subsection{Bound trajectory}

Having derived the position operator in Eq. \eqref{ZB1}, we now trace out the electron trajectory in the system. In general, a free electron can travel in a region as large as the area of the system defined by its physical boundaries, e.g., edges or interfaces. However, we show that in the TI thin film system, the electron trajectory is bound within an area determined by the initial state (spin and momentum) of the electron and the energy gap of the system. Consider an electron injected into the top surface of the TI film with initial spin state in the spin up direction and momentum $\bm{k}$ which is represented by the planewave $|\psi_0({\bm k})\rangle=e^{i{\bm k}\cdot {\bm r}}|\phi_0\rangle$. The position of the electron on the $x$-$y$ plane at time $t$ can be calculated from Eq. \eqref{ZB1} and is given explicitly by
\begin{eqnarray}\label{RZB}
x({\bm k},t)=&&X_- (\cos w_- t-1)-X_+(\cos w_+ t-1),\\
y({\bm k},t)=&&Y_-(\cos w_- t-1)-Y_+(\cos w_+ t-1)\nonumber\\
&&+Y_0( \cos w_2 t - \cos w_1 t),
\end{eqnarray}
where $X_\pm=v_f^2\frac{(1\mp\sin\Theta)(k_u\pm k_v\sin\Phi)}{w_\pm^2}$,
$Y_\pm=v_f^2\frac{(1\mp\sin\Theta)k_v\cos \Phi\sin\Theta}{w_\pm^2}$, and
$Y_0=v_f^2 \frac{2\cos^2\Theta k_v\cos \Phi}{w_-w_+}$.

Corresponding expressions for other combinations of injected spin orientation and injection surfaces can be obtained  from symmetry arguments. Eq. \eqref{Hal} in terms of $k_\Delta$ and $k_B$ is, explicitly,
\begin{equation}
  H/v_f = \tau_z (k_y\sigma_x - \sigma_y k_x) + k_\Delta\tau_x – k_B\sigma_x \label{Hal1}. 
\end{equation}
Eq. \eqref{Hal1} is invariant upon a simultaneous $\tau$ reflection about the $\tau_x$ axis and in-plane spatial inversion, i.e. $\tau_z \rightarrow -\tau_z, x \rightarrow -x, y \rightarrow -y$. This implies that the $x$ and $y$ displacements of electrons injected into the top and bottom surfaces have the same magnitudes but opposite signs. 
Eq. \eqref{Hal1} is also invariant upon a simultaneous spin reflection about $\sigma_x$  ($\sigma_{y,z} \rightarrow \sigma_{y,z}$) and reflection along the $y$ axis ($x \rightarrow -x$,$y \rightarrow y$). This implies that spin up and spin down electrons injected into a given surface (top / bottom) have the same $x$ displacements, and $y$ displacements of the same magnitude but opposite signs.

\begin{figure}
\includegraphics[width=0.5\textwidth]{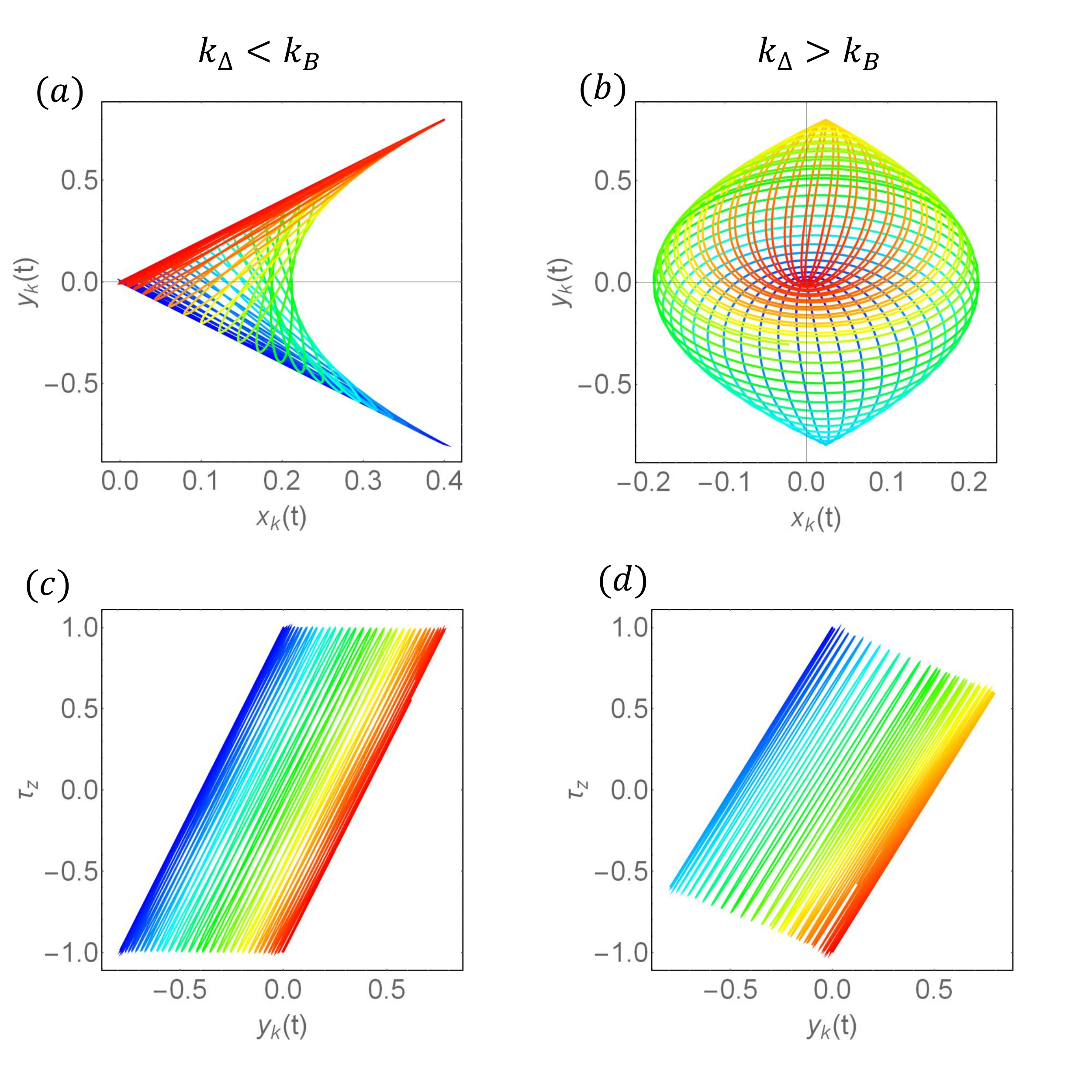}
\caption{(a)--(b) Trajectory of an electron with up-spin on the $x-y$ plane for different ratios of $k_\Delta/k_B$. (c)--(d) Projection of trajectory on the $y-\tau_z$ plane, in which the pseudo-spin index $\tau_z(t)\equiv\langle\tau_z(t)\rangle $ indicates whether the electron is on the top or bottom surfaces.}
\label{Fig2}
\end{figure}

The electron motion of an electron injected in the top surface with initial spin in the +z direction on the $x-y$ plane is depicted in Fig. \ref{Fig2}(a) and (b) for different ratios of $k_\Delta/k_B$. Taking the initial position of the electron to be the origin, it can be shown that $x(t)\ge 0$ for $k_\Delta/k_B<1$, i.e., the electron is always confined in the $+x$-half of the $x$-$y$ plane. On the other hand, when $k_\Delta/k_B>1$, the trajectory of the injected electron encompasses the origin as shown in Fig. \ref{Fig2}(b). 

It can be seen that the electron oscillation comprises both transverse and longitudinal modes. This is a manifestation of the four-band system illustrated in Fig. \ref{Fig1}(b), where the quantum dynamics involves not just the evolution of the spin, but also the pseudo-spin degree of freedom, which in our case, represents the surface index (top and bottom surfaces). The electron trajectories in Fig. \ref{Fig2}(a) and (b) indicate the presence of oscillations in both the transverse ($y$) and longitudinal ($x$) directions. Now in the conventional ZB picture, an  electron injected along the $x$-direction would undergo oscillations in the transverse $y$-direction, due to the electron spin precession and spin-momentum locking. In this simple picture, the longitudinal oscillations do not seem to play a role. To explain the emergence of the longitudinal oscillations, we need to consider the pseudo-spin ($\tau_z$) degree of freedom. This can be ascribed to the precession of the pseudo-spin, which represents the back and forth tunneling between surfaces. From Eq. \eqref{Hal}, this pseudo-spin dynamics is coupled to the longitudinal motion. Indeed, as shown in Fig. \ref{Fig2}(d), the electron lies in the positive $x$-half when it is on the top surface, and would move to the negative $x$-half after tunneling to the bottom surface. Thus, the back and forth tunneling between the surfaces mediated by the hybridization $\Delta$ translates into the oscillation of the electron motion in the longitudinal $x-$direction.

In the thick TI film limit where the top and bottom surfaces are decoupled, i.e., $\Delta=0$ in the Hamiltonian Eq. \eqref{Hal}, the motion of the electron is simply given by ${\bm r}(t)=\frac{1}{k^2}(z\times {\bm k}) \left [1-2\cos^2{kv_f t}\right]$. Surprisingly, a spin-up electron initially injected along the $x$-direction will only move in the $y$-direction, i.e, its trajectory is confined in a line perpendicular to the injection direction. This can be explained by considering the electron velocity given by ${\bm v}=\partial_{\bm k}H=v_f(z\times {\bm \sigma})$. The electron spin precesses as ${\bm \sigma}={\bm k}\sin{2kv_f t}$, yielding ${\bm v}=\frac{v_f}{k}(z\times {\bm k})\sin{2kv_f t}$, which is perpendicular to the momentum.

\subsection{Wavepacket dynamics}
In the previous section, we have considered the trajectory of a single electron. We now consider the more practical case of an electron wavepacket, which is a superposition of different momentum states. In general, the beat frequencies $w$s as given in Eq. \eqref{freq} are dependent on the momentum. Thus, when evaluating the expectation value of the position operator for a wavepacket, the resulting interference of oscillations with different momentum-dependent frequencies would, in general, lead to a decay of the ZB over time. In order to sustain the ZB motion, we need to realize a scenario where at least one beat frequency is momentum-independent.  We will show that such a scenario can be achieved by the appropriate choice of parameters such as hybridization energy and the in-plane magnetic field. 

Suppose that the electron is injected in the $y$-direction, i.e., $k_x=0,k_y=k$, and the hybridization and magnetic field are tuned so that $k_\Delta=k_B$. In this case, the beat frequencies of Eq. \eqref{freq} are now given by
\begin{eqnarray}\label{freq2}
w_1=	2v_fk_u, w_2=2v_fk_B, w_\pm=2v_f(k_u\pm k_B), \label{w1}
\end{eqnarray}
in which we recall that $k_u=\sqrt{k_\Delta^2+k_y^2}$.

We can see that besides the three momentum-dependent frequencies, there is one frequency $w_2$ that is independent of momentum. At large time scales, we would expect the oscillations associated with the other three frequencies to decay away due to interference, while the oscillations associated with the $k$-independent $w_2$ frequency would persist.  This is one of the main results of this paper.

\begin{figure}
\includegraphics[width=0.5\textwidth]{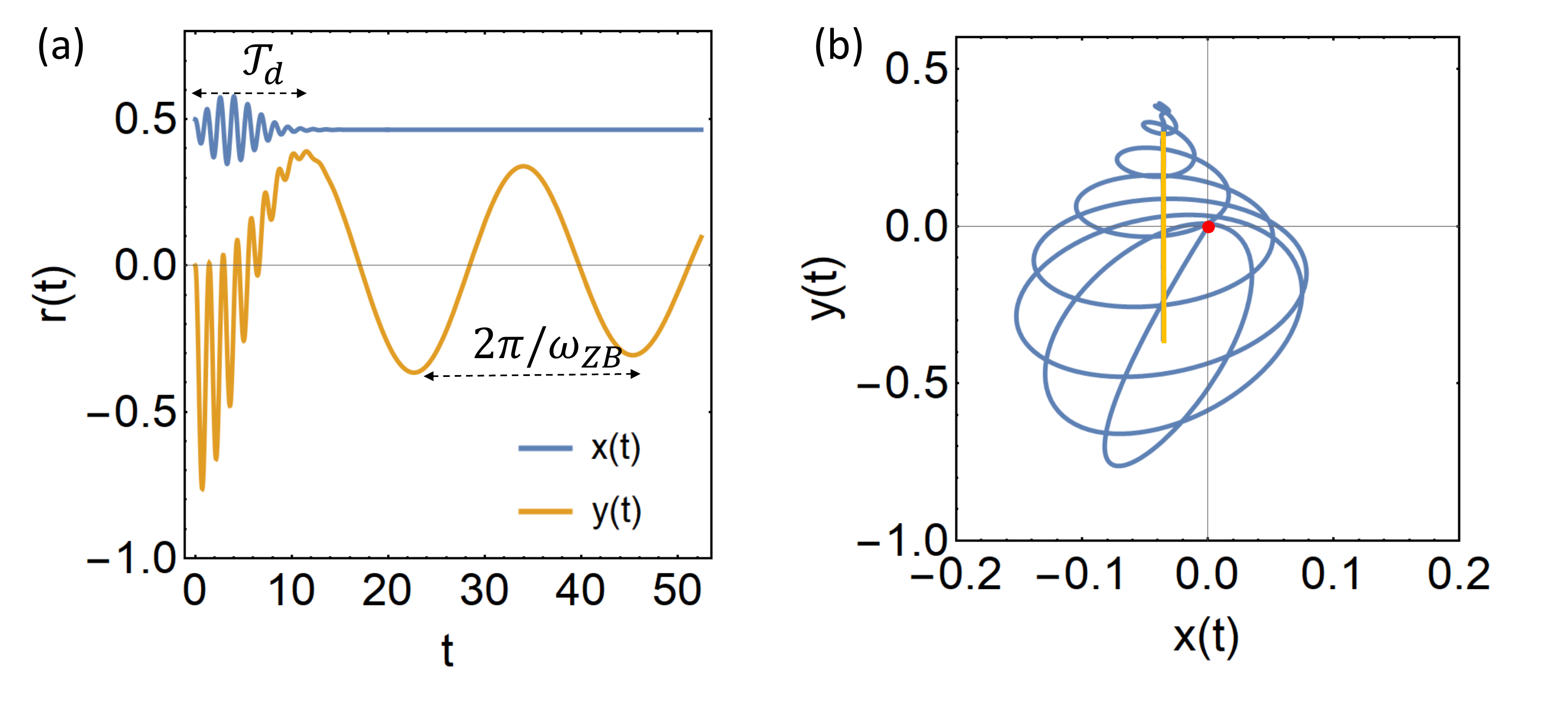}
\caption{(a)  The position of the electron wavepacket as  a functions of time and (b) oscillatory trajectory on the $x$-$y$ plane. In the special case $k_B=k_\Delta$, there is a persistent transverse oscillation mode with frequency $w_{ZB}=2\Delta$, whereas all the other modes decay after a time interval $\mathcal{T}_d$. In (b), the initial position is marked by a red dot, while the final steady-state transverse oscillation is marked by an orange line.}
\label{Fig3}
\end{figure}

To quantitatively verify the above intuitive picture of persistent ZB motion, we consider the electron wavepacket given by
\begin{equation}
|\psi_0({\bm k})\rangle=a({\bm k})|\phi_0\rangle,
\end{equation}
where $|\phi_0\rangle$ is the initial spin state, and $a({\bm k})=\frac{1}{\sqrt{\pi}\delta k}e^{-\frac{({\bm k}-{\bm k}_0)^2}{2\delta k^2}}$ is the Gaussian distribution function that represents the spread of the electron state in momentum space, in which ${\bm k}_0$ and $\delta k$ are the initial momentum and line-width, respectively.  The expectation value of the position operator Eq. \eqref{ZB1} for the above state is given by
\begin{eqnarray}\label{rfull}
{\bm r}_Z(t)=\int d{\bm k} |a({\bm k})|^2\langle\phi_0|{\hat{\bm r}}_Z({\bm k},t)|\phi_0\rangle, \label{e14}
\end{eqnarray}
where the integration is taken over momentum space. As a consequence of the wavepacket spread in $k$-space, the ZB will generally decay over time. In order to analytically describe the damping process, we will consider the narrow wavepacket limit, i.e., $\delta k/k_0 \ll 1$, so that the integration of the Gaussian function in Eq. \eqref{rfull} can be approximated by
\begin{eqnarray}\label{approx}
{\bm r}_Z(t)\approx {\bm r}_Z({\bm k}_0,t)+\frac{\delta k^2}{4} \nabla_{\bm k}^2 {\bm r}_Z({\bm k}_0,t) \label{e15}
\end{eqnarray}
up to $\mathcal{O}(\delta k^4)$.
In the above, the first term is the initial ZB oscillation with momentum ${\bm k}_0$, and the second term represents the deviation of the ZB around the packet center. Substituting the position operator in Eq. \eqref{ZB1}, we have
\begin{eqnarray}
\partial_{k_a}^2 {\bm r}(k,t)=&&\sum_{ij}e^{i\Omega_{ij}t}\partial_{k_a}^2 {\bm A}_{ij}-e^{i\Omega_{ij}t}{\bm A}_{ij} (\partial_{k_a}\Omega_{ij})^2t^2 \nonumber\\
&&+2e^{i(\Omega_{ij}t+\pi/2)}\partial_{k_a}{\bm A}_{ij} (\partial_{k_a}\Omega_{ij})t, \label{e16}
\end{eqnarray}
in which $a=x,y$. The first term in the above describes oscillations with constant amplitude that are in-phase with the initial oscillation. The next two terms have time-dependent amplitudes that are linear and quadratic in time, respectively. Rearranging the equation (\ref{approx}), the ZB of a wavepacket can be expressed as
\begin{eqnarray}\label{ZB3}
{\bm r}_{Z}(t)=&&\sum_{i\ne j} {{\bm r}_{ij}}({\bm k}_0,t)\left(1-\frac{t^2}{{\mathcal T}_{ij}^2}\right)
\end{eqnarray}
where the decay time is defined as
\begin{eqnarray}
{\mathcal T}_{ij}={\frac{2}{\delta k|\nabla_{\bm k}\Omega_{ij}({\bm k}_0)|}}
\end{eqnarray}
with the beat frequencies given by Eq. \eqref{freq}. At short $t$, the first term in Eq. \eqref{ZB3} can be formally written as ${\bm r}_{Z}(t)\approx\sum_{ij} {{\bm r}}_{ij}({\bm k}_0,t) e^{-t^2/T_{ij}^2}$, which expresses the exponential decay of the ZB (see Fig. \ref{Fig3}).

From Eq. \eqref{freq2}, the decay times are obtained as
	\begin{eqnarray}
	\frac{1}{{\mathcal T}_2}&&=\theta(k_\Delta^2-k_B^2)\frac{v_f\delta k k_0}{\sqrt{k_\Delta^2+k_0^2}},\label{T2}\\
	\frac{1}{{\mathcal T}_d}&&=\frac{v_f\delta k k_0}{\sqrt{k_\Delta^2+k_0^2}},\label{Td}
	\end{eqnarray}
where $\theta(x)$ is the Heaviside step function, ${\mathcal T}_2$ corresponds to the combinations of $i$ and $j$ where $|\Omega_{ij}|=w_2$, and ${\mathcal T}_d$  the other combinations of $i$ and $j$.  As can be seen, when one of the beat frequencies, i.e., $w_2$, becomes independent of momentum at resonance where  $k_{\Delta}=k_B$, the associated decay time ${\mathcal T}_2$ in Eq. \eqref{T2} goes to infinity. This implies that the ZB related to this mode will be persistent.  In this case, the steady state transverse oscillation is given by 
\begin{eqnarray}
y(t)=Y_0(k_0)\cos(2v_fk_\Delta t).
\end{eqnarray}

In the limit of large hybridization $k_\Delta\gg k_0$, the persistent oscillation reduces to $y(t)\approx A_{ZB}\cos(\omega_{ZB}t)$ with $A_{ZB}=\frac{1}{2k_\Delta}$ and $\omega_{ZB}=2v_fk_\Delta=2\Delta$ being respectively the amplitude and frequency. This persistent oscillation is depicted by the orange line in Fig. \ref{Fig3}(b).  Surprisingly, both the amplitude and frequency of the persistent mode do not depend on the initial momentum and width of the injected wavepacket and are instead determined by a single parameter, i.e., the hybridization energy. Following Eq. \eqref{T2} , the ZB has a sharp transition from a transient to persistent mode at $k_\Delta=k_B$ at which the bulk gap closes (Fig. \ref{Fig1}) and the TI film undergoes a topological phase transition \cite{zyuzin2011parallel}. We can hence refer to the persistent oscillation as a topological mode of electron oscillation.

\section{Electron-mediated RKKY interaction}\label{sec3}
In the previous section, we have shown that the electron trajectory is confined and may be highly anisotropic (see e.g., Fig. \ref{Fig2}(a)).  This has consequences for the transport-related properties of the system, such as the electron-mediated RKKY interaction. The confinement of the electron trajectory implies that the electrons are not able to mediate information, e.g. angular momentum, between magnetic moments separated by a separation distance that exceeds the confinement region. In order to verify this effect, we consider two magnetic centers $S_i$ ($i=1,2$) located at $\bm {R}_i$. The electron-mediated exchange interaction between the magnetic centers is modeled by 
\begin{eqnarray}\label{Hex}
H_{int}=J\sum_{i=1,2}{\bm S}_i\cdot{\bm \sigma}(\bm{r}-\bm{R}_i),
\end{eqnarray}
where $J$ is the exchange coupling. The exchange interaction can be considered as a perturbation to the Hamiltonian in Eq. \eqref{Hal}. For simplicity, we assume that $\bm {R}_1=(0,0)$, and $\bm {R}_2=R(\cos\phi_R,\sin\phi_R)$. We show that the RKKY coupling between the two magnetic centers does not depend on just the distance $R$, but also on the direction $\phi_R$ between them.

In the framework of the second-order perturbation theory, the effective interaction between two magnetic impurities is given by \cite{RKKY1,RKKY2,RKKY3,mattis2006theory,PhysRevB.54.12953,PhysRevB.81.113302,RKKYTI:PRL2011}
\begin{eqnarray}
&&H_{RKKY}=-\frac{J^2}{\pi}\mathrm{Tr}\int_{-\infty}^{\epsilon_F}d\epsilon\nonumber\\
&&\times\mathrm{Im}\left[({\bm S}_1\cdot{\tilde {\bm \sigma}})G({\bm R},\epsilon^+)({\bm S}_2\cdot{\tilde {\bm \sigma}})G(-{\bm R},\epsilon^+)\right],
\end{eqnarray}
where $\epsilon^+=\epsilon+i0^+$, Tr stands for the trace over the spin degree of freedom, and the expanded spin operator in spin and pseudo-spin spaces is defined as ${\tilde {\bm \sigma}}=\tau_0\otimes{\bm \sigma}$, in which $\tau_0$ is the identity matrix of rank 2. The Green's function in real space is given by the Fourier transformation
\begin{eqnarray}
G({\bm R},\epsilon^+)=\int \frac{d^2 {\bm q}}{A_{BZ}}e^{i{\bm q}\cdot {\bm R}}G({\bm q},\epsilon^+),
\end{eqnarray}
where $G({\bm q},\epsilon^+)=[\epsilon^+-H_0({\bm q})]^{-1}$ is the Green's function in momentum space, and $A_{BZ}$ is the area of the first Brillouin zone.

Let us first consider the weak hybridization limit, i.e., $k_\Delta \ll k_B$. In this limit, the system is gapless and the two Dirac points are separated by ${\bm q}_0\approx {\bm k}_B$. The analytical expression of the RKKY coupling can be obtained as (see Appendix \ref{App2} for more details)
\begin{eqnarray}\label{Hweak}
H_{RKKY}^{\mathrm{weak}}=F_1({\bm S}_1\cdot {\bm S}_2)&&+F_2({\bm S}_1\cdot {\hat{\bm u}})({\bm S}_2 \cdot {\hat{\bm u}})\nonumber\\
&&+F_3\left[{\hat{\bm u}}\cdot({\bm S}_1\times {\bm S}_2)\right],
\end{eqnarray}
in which the range functions are
\begin{eqnarray}
F_1&&=- \cos(2{\bm R}\cdot {\bm q}_0)\frac{64\pi J^2\epsilon_f}{v_f^2A_{BZ}^2R^2} \sin\frac{2R\epsilon_f}{v_f},\\
F_2&&= \cos({2\bm R}\cdot {\bm q}_0)\frac{64\pi J^2\epsilon_f}{v_f^2A_{BZ}^2R^2} \sin\frac{2R\epsilon_f}{v_f},\\
F_3&&=-\sin({2\bm R}\cdot {\bm q}_0)\frac{64\pi J^2\epsilon_f}{v_f^2A_{BZ}^2R^2} \sin\frac{2R\epsilon_f}{v_f}.
\end{eqnarray}
 In the above, ${\hat{\bm u}}=({\hat{\bm R}}\times {\bm z})$, with ${\hat{\bm R}}={{\bm R}}/R$ being the unit vector along ${\bm R}$.

The RKKY coupling in Eq. \eqref{Hweak} consists of three terms: the Heisenberg exchange, the spin-frustrated, and the Dzyaloshinsky-Moriya interaction terms. As shown above, the RKKY coupling exhibits not only the usual $R^{-2}$ distance dependence \cite{RKKYTI:PRL2011}  in a semi-infinite thick TI slab with only a single surface, but also has an additional direction-dependence due to the $\cos(2{\bm q}_0\cdot {\bm R})$ factor that is absent in the semi-infinite thick slab. This directional dependence  stems from the contribution of the surface states on both surfaces of the film in mediating the effective exchange coupling, and the fact that the corresponding Dirac cones are separated in momentum space. In the case where the two magnetic impurities are separated along the $x$-direction, i.e., along the in-plane magnetic field direction, ${\bm k}_B\cdot {\bm R}=0$ and the RKKY coupling reaches its maximum. 

This can be explained by considering the process of indirect exchange coupling between the two magnetic moments via the itinerant electrons. When an electron is in close proximity to the first magnetic moment, its spin angular momentum is coupled to that of the magnetic moment. If there is finite electron overlap with the second magnetic moment, then its spin angular momentum is also coupled to the second moment. In this way, an effective exchange coupling arises between the two magnetic moments. The strength of the effective coupling depends on the rate and probability of electron overlap between one magnetic moment and the other. In other words, if the second magnetic moment is located at a position with little electron overlap with the first magnetic moment, then the coupling between the moments would be weak. Conversely, if the second magnetic moment is at a position where the electron has a high probability of overlap, the coupling will be enhanced. In our case, when the magnetic field is applied along the $x$-direction, the electron motion has a tendency of being confined along the same $x$-direction [see Fig. \ref{Fig2}(a)]. This means that a second magnetic moment placed along the $x$-direction with respect to the first moment will have high probability of being coupled by an intermediary electron, thus inducing stronger RKKY coupling.

Although Fig. \ref{Fig1} shows only the results of a spin up electron injected on the top surface explicitly, the results of the symmetry analysis following Eq. \eqref{Hal1} imply that the electron trajectory will still be confined along the $x$ direction for spins of other orientations injected into both the top and the bottom surfaces. 

To quantify the correlation between the RKKY coupling and the electron trajectory, we will analyze the preferred direction of the electron motion. As the electron position oscillates over time as described in Eq. \eqref{RZB}, we consider its average value ${\bar {\bm r}}(k)=\lim_{T\rightarrow\infty}\frac{1}{T}\int_0^T r(t)$, which is explicitly given by
\begin{eqnarray}\label{average}
{\bar x}(k)=-\frac{((k_u^2-k_v^2)k_v\sin\Phi+k_u(k_u^2+k_v^2\cos 2\Phi)\sin\Theta)}{2(k_u^4+k_v^4+2k_u^2k_v^2\cos 2\Phi)},\nonumber\\
{\bar y}(k)=-\frac{k_v\cos\Phi \sin\Theta(2k_uk_v\sin\Phi+(k_u^2+k_v^2)\sin\Theta)}{2(k_u^4+k_v^4+2k_u^2k_v^2\cos 2\Phi)}.\nonumber\\
\end{eqnarray}
In the limit of weak hybridization energy $k_\Delta\ll k_B$, the above reduce to
\begin{eqnarray}\label{rweak}
{\bar {\bm r}(k)}=\frac{z\times ({\bm k}-{\bm k}_B)}{2|({\bm k}-{\bm k}_B)|^2}. 
\end{eqnarray}
Eq. \eqref{rweak} is the time-averaged position of an electron with momentum $\bm k$. Now, averaging the above over momentum space up to the Fermi wave-vector, we obtain
\begin{eqnarray}
{\bar {\bm r}}=-\frac{1}{2k_f^2}(z\times {\bm k}_B), \label{e31}
\end{eqnarray}
which indicates that the electron will preferably move in the direction perpendicular to the direction separating the two Dirac cones. Therefore, when ${\bm R}$ is parallel to ${\bar {\bm r}}$, and thus perpendicular to $\bm{k}_B$, the RKKY coupling strength will be maximum. This is in line with the prediction based on the electron trajectory, as discussed above.

We note that in the above, the preferred motion direction was obtained based on the position of the plane-wave electron. Here, we show that the preferred direction is the same if we consider the electron wavepacket treatment. As the Gaussian function in the wavepacket picture is time-independent, it would not alter the position value after time-averaging. From Eq. \eqref{e14}, the average position of an electron wavepacket initially centered at $k_0$ is simply derived as ${\bar {\bm r}}_{pk}(k_0)=\sum_k |a(k-k_0)|^2{\bar {\bm r}(k)}\approx {\bar {\bm r}(k_0)}+\delta{\bar {\bm r}(k_0)}$, where the deviation $\delta{\bar {\bm r}(k_0)}=\frac{\delta k^2}{4}\nabla_k^2 {\bar {\bm r}(k_0)}$  follows Eq. \eqref{e15} for a narrow wavepacket, with ${\bar {\bm r}(k)}$ given in Eq. \eqref{average}. In the weak hybridization limit, applying Eq. \eqref{rweak}, we find that the deviation $\delta{\bar {\bm r}(k_0)}=0$, which means that the preferred direction of motion of a wavepacket coincides with that of a plane-wave electron. This result thus suggests that one may use the wavepacket treatment in understanding properties of the RKKY coupling, besides the conventional plane Bloch wave approaches \cite{RKKY1,RKKY2,RKKY3} in future works.

 On the other hand, in the strong hybridization limit $k_\Delta\gg k_B$, the RKKY coupling is given by (details are shown in Appendix \ref{App2})
\begin{eqnarray}
H_{RKKY}^{\mathrm{strong}}={\tilde F}_1({\bm S}_1\cdot {\bm S}_2)+&&{\tilde F}_2({\bm S}_1\cdot {\hat{\bm u}})({\bm S}_2 \cdot {\hat{\bm u}})\nonumber\\
+&&{\tilde F}_3\left[{\hat{\bm u}}\cdot({\bm S}_1\times {\bm S}_2)\right],
\end{eqnarray}
where the range functions are now given by
\begin{eqnarray}
{\tilde F}_1&&=-\frac{16\pi J^2}{v_f^4A_{BZ}^2} \mathrm{Im}\int_{-\infty}^{\epsilon_f} d\epsilon({\tilde A}_1^2+{\tilde A}_2^2-{\tilde B}^2),\\
{\tilde F}_2&&=-\frac{32\pi J^2}{v_f^4A_{BZ}^2}\mathrm{Im}\int_{-\infty}^{\epsilon_f}d\epsilon {\tilde B}^2,\\
{\tilde F}_3&&=\frac{32\pi J^2}{v_f^4A_{BZ}^2}\mathrm{Im}\int_{-\infty}^{\epsilon_f} d\epsilon {\tilde A}_1{\tilde B},
\end{eqnarray}
with ${\tilde A}_1=\epsilon K_0(\frac{R|w|}{v_f}), {\tilde A}_2=\Delta K_0(\frac{R|w|}{v_f}), {\tilde B}=|w|K_1(\frac{R|w|}{v_f})$, and where $w=\sqrt{\Delta^2-\epsilon^2}$.

In this case, the surface states are gapped (see Fig. \ref{Fig1}(d)), and the Dirac cones vanish. In this limit, the RKKY coupling becomes isotropic, i.e., it is independent of the angle between the magnetic centers. This result is consistent with the calculated electron trajectory in the gapped scenario, where its trajectory is almost isotropic in the 2D plane (see Fig. \ref{Fig2}(c)). This can be further verified by considering the time-averaged electron position as outlined above, which is given by
\begin{eqnarray}
{\bar {\bm r}_k}=\frac{1}{k_u^2}\left(k_y,-\frac{k_y^2}{k_\Delta^2}k_x\right).
\end{eqnarray}
The above goes to zero upon averaging over momentum space, so that there is no preferred direction of the electron motion in the 2D plane in this case.

We remark here that in the insulating phase, the TI film has been shown to have a diamagnetic response to an in-plane magnetic field \cite{zyuzin2011parallel}. As a consequence, the magnetic moments in the TI film may acquire an additional magnetic response and the steady state magnetization may change accordingly. However, the magnetic susceptibility is extremely small, i.e. on the order of $10^{-8}$ \cite{zyuzin2011parallel}, which is several orders of magnitude smaller than even the small diamagnetic susceptibility of typical metals. The effect of the induced magnetization can thus be neglected in the bandstructure of the TI film. Since the RKKY coupling is derived from the bandstructure of the TI film, therefore it will not susceptible to this diamagnetic response.  

In addition, we note that if the Fermi level lies within the gap in the insulating phase, the RKKY mechanism is no longer be valid as it relies on itinerant electrons. Instead, the indirect exchange coupling is now described by the van Vleck mechanism as discussed in previous works \cite{vanVleck,vanVleck2,vanVleck3}. In our work, we assume that the Fermi level is finite, i.e., within the conduction band, and ignore the van Vleck coupling for simplicity. 

\section{Conclusion}\label{sec4}
In this paper, we investigated the anomalous motion of electrons in topological insulator thin films. First, we showed that due to the hybridization of the surface states with opposite helicities, a spin-polarized electron will undergo oscillatory motion within a confined region. Furthermore, the oscillation is anisotropic with the preferred direction being along the separation of the two Dirac points, a finding which that be ascribed to the anisotropy of the Fermi circle. 
As a consequence, the direction and distance dependence of RKKY interactions mediated by
itinerant electrons between two magnetic impurities in thin TI films have a strong correlation with the electron
motion. Interestingly, it was found that the RKKY coupling is maximized when two impurities at a fixed distance are positioned along the separation direction of the two Dirac
points. This finding is consistent with the preferred direction of the confined electron motion. 

\section*{Acknowledgments}
We would like to acknowledge the following for funding support: MOE Tier I (NUS Grant No. R-263-000-D66-114), MOE Tier II MOE2018-T2-2-117 (NUS Grant No. R-398-000-092-112), MOE RSB and CFS funds (NUS Grant Nos. C-261-000-207-532 and C-261-000-777-532), and NRF-CRP12-2013-01 (NUS Grant No. R-263-000-B30-281).


\begin{thebibliography}{71}%
	\makeatletter
	\providecommand \@ifxundefined [1]{%
		\@ifx{#1\undefined}
	}%
	\providecommand \@ifnum [1]{%
		\ifnum #1\expandafter \@firstoftwo
		\else \expandafter \@secondoftwo
		\fi
	}%
	\providecommand \@ifx [1]{%
		\ifx #1\expandafter \@firstoftwo
		\else \expandafter \@secondoftwo
		\fi
	}%
	\providecommand \natexlab [1]{#1}%
	\providecommand \enquote  [1]{``#1''}%
	\providecommand \bibnamefont  [1]{#1}%
	\providecommand \bibfnamefont [1]{#1}%
	\providecommand \citenamefont [1]{#1}%
	\providecommand \href@noop [0]{\@secondoftwo}%
	\providecommand \href [0]{\begingroup \@sanitize@url \@href}%
	\providecommand \@href[1]{\@@startlink{#1}\@@href}%
	\providecommand \@@href[1]{\endgroup#1\@@endlink}%
	\providecommand \@sanitize@url [0]{\catcode `\\12\catcode `\$12\catcode
		`\&12\catcode `\#12\catcode `\^12\catcode `\_12\catcode `\%12\relax}%
	\providecommand \@@startlink[1]{}%
	\providecommand \@@endlink[0]{}%
	\providecommand \url  [0]{\begingroup\@sanitize@url \@url }%
	\providecommand \@url [1]{\endgroup\@href {#1}{\urlprefix }}%
	\providecommand \urlprefix  [0]{URL }%
	\providecommand \Eprint [0]{\href }%
	\providecommand \doibase [0]{http://dx.doi.org/}%
	\providecommand \selectlanguage [0]{\@gobble}%
	\providecommand \bibinfo  [0]{\@secondoftwo}%
	\providecommand \bibfield  [0]{\@secondoftwo}%
	\providecommand \translation [1]{[#1]}%
	\providecommand \BibitemOpen [0]{}%
	\providecommand \bibitemStop [0]{}%
	\providecommand \bibitemNoStop [0]{.\EOS\space}%
	\providecommand \EOS [0]{\spacefactor3000\relax}%
	\providecommand \BibitemShut  [1]{\csname bibitem#1\endcsname}%
	\let\auto@bib@innerbib\@empty
	\bibitem [{\citenamefont {Schr{\"o}dinger}(1930)}]{schrodinger1930kraftefreie}%
	\BibitemOpen
	\bibfield  {author} {\bibinfo {author} {\bibfnamefont {E.}~\bibnamefont
			{Schr{\"o}dinger}},\ }\href@noop {} {\bibfield  {journal} {\bibinfo
			{journal} {Preuss. Akad. Wiss. Phys. Math. KI.}\ }\textbf {\bibinfo {volume}
			{24}},\ \bibinfo {pages} {418} (\bibinfo {year} {1930})}\BibitemShut
	{NoStop}%
	\bibitem [{\citenamefont {J.~Schliemann}\ and\ \citenamefont
		{Westervelt}(2005)}]{Schli05}%
	\BibitemOpen
	\bibfield  {author} {\bibinfo {author} {\bibfnamefont {D.~L.}\ \bibnamefont
			{J.~Schliemann}}\ and\ \bibinfo {author} {\bibfnamefont {R.~M.}\ \bibnamefont
			{Westervelt}},\ }\href@noop {} {\bibfield  {journal} {\bibinfo  {journal}
			{Phys. Rev. Lett.}\ }\textbf {\bibinfo {volume} {94}},\ \bibinfo {pages}
		{206801} (\bibinfo {year} {2005})}\BibitemShut {NoStop}%
	\bibitem [{\citenamefont {Schliemann}\ \emph {et~al.}(2006)\citenamefont
		{Schliemann}, \citenamefont {Loss},\ and\ \citenamefont
		{Westervelt}}]{schliemann2006zitterbewegung}%
	\BibitemOpen
	\bibfield  {author} {\bibinfo {author} {\bibfnamefont {J.}~\bibnamefont
			{Schliemann}}, \bibinfo {author} {\bibfnamefont {D.}~\bibnamefont {Loss}}, \
		and\ \bibinfo {author} {\bibfnamefont {R.}~\bibnamefont {Westervelt}},\
	}\href@noop {} {\bibfield  {journal} {\bibinfo  {journal} {Physical Review
				B}\ }\textbf {\bibinfo {volume} {73}},\ \bibinfo {pages} {085323} (\bibinfo
		{year} {2006})}\BibitemShut {NoStop}%
	\bibitem [{\citenamefont {Demikhovskii}\ \emph {et~al.}(2008)\citenamefont
		{Demikhovskii}, \citenamefont {Maksimova},\ and\ \citenamefont
		{Frolova}}]{demikhovskii2008wave}%
	\BibitemOpen
	\bibfield  {author} {\bibinfo {author} {\bibfnamefont {V.~Y.}\ \bibnamefont
			{Demikhovskii}}, \bibinfo {author} {\bibfnamefont {G.}~\bibnamefont
			{Maksimova}}, \ and\ \bibinfo {author} {\bibfnamefont {E.}~\bibnamefont
			{Frolova}},\ }\href@noop {} {\bibfield  {journal} {\bibinfo  {journal}
			{Physical Review B}\ }\textbf {\bibinfo {volume} {78}},\ \bibinfo {pages}
		{115401} (\bibinfo {year} {2008})}\BibitemShut {NoStop}%
	\bibitem [{\citenamefont {Biswas}\ and\ \citenamefont
		{Ghosh}(2012)}]{biswas2012zitterbewegung}%
	\BibitemOpen
	\bibfield  {author} {\bibinfo {author} {\bibfnamefont {T.}~\bibnamefont
			{Biswas}}\ and\ \bibinfo {author} {\bibfnamefont {T.~K.}\ \bibnamefont
			{Ghosh}},\ }\href@noop {} {\bibfield  {journal} {\bibinfo  {journal} {Journal
				of Physics: Condensed Matter}\ }\textbf {\bibinfo {volume} {24}},\ \bibinfo
		{pages} {185304} (\bibinfo {year} {2012})}\BibitemShut {NoStop}%
	\bibitem [{\citenamefont {Biswas}\ and\ \citenamefont
		{Ghosh}(2014)}]{biswas2014wave}%
	\BibitemOpen
	\bibfield  {author} {\bibinfo {author} {\bibfnamefont {T.}~\bibnamefont
			{Biswas}}\ and\ \bibinfo {author} {\bibfnamefont {T.~K.}\ \bibnamefont
			{Ghosh}},\ }\href@noop {} {\bibfield  {journal} {\bibinfo  {journal} {Journal
				of Applied Physics}\ }\textbf {\bibinfo {volume} {115}},\ \bibinfo {pages}
		{213701} (\bibinfo {year} {2014})}\BibitemShut {NoStop}%
	\bibitem [{\citenamefont {Shi}\ \emph {et~al.}(2013)\citenamefont {Shi},
		\citenamefont {Zhang},\ and\ \citenamefont {Chang}}]{Shi13}%
	\BibitemOpen
	\bibfield  {author} {\bibinfo {author} {\bibfnamefont {L.-k.}\ \bibnamefont
			{Shi}}, \bibinfo {author} {\bibfnamefont {S.-c.}\ \bibnamefont {Zhang}}, \
		and\ \bibinfo {author} {\bibfnamefont {K.}~\bibnamefont {Chang}},\
	}\href@noop {} {\bibfield  {journal} {\bibinfo  {journal} {Phys. Rev. B}\
		}\textbf {\bibinfo {volume} {87}},\ \bibinfo {pages} {161115} (\bibinfo
		{year} {2013})}\BibitemShut {NoStop}%
	\bibitem [{\citenamefont {Rusin}\ and\ \citenamefont
		{Zawadzki}(2007)}]{Rusin07}%
	\BibitemOpen
	\bibfield  {author} {\bibinfo {author} {\bibfnamefont {T.~M.}\ \bibnamefont
			{Rusin}}\ and\ \bibinfo {author} {\bibfnamefont {W.}~\bibnamefont
			{Zawadzki}},\ }\href@noop {} {\bibfield  {journal} {\bibinfo  {journal}
			{Phys. Rev. B}\ }\textbf {\bibinfo {volume} {76}},\ \bibinfo {pages} {195439}
		(\bibinfo {year} {2007})}\BibitemShut {NoStop}%
	\bibitem [{\citenamefont {Rusin}\ and\ \citenamefont
		{Zawadzki}(2008)}]{PhysRevB.78.125419}%
	\BibitemOpen
	\bibfield  {author} {\bibinfo {author} {\bibfnamefont {T.~M.}\ \bibnamefont
			{Rusin}}\ and\ \bibinfo {author} {\bibfnamefont {W.}~\bibnamefont
			{Zawadzki}},\ }\href {\doibase 10.1103/PhysRevB.78.125419} {\bibfield
		{journal} {\bibinfo  {journal} {Phys. Rev. B}\ }\textbf {\bibinfo {volume}
			{78}},\ \bibinfo {pages} {125419} (\bibinfo {year} {2008})}\BibitemShut
	{NoStop}%
	\bibitem [{\citenamefont {Maksimova}\ \emph {et~al.}(2008)\citenamefont
		{Maksimova}, \citenamefont {Demikhovskii},\ and\ \citenamefont
		{Frolova}}]{PhysRevB.78.235321}%
	\BibitemOpen
	\bibfield  {author} {\bibinfo {author} {\bibfnamefont {G.~M.}\ \bibnamefont
			{Maksimova}}, \bibinfo {author} {\bibfnamefont {V.~Y.}\ \bibnamefont
			{Demikhovskii}}, \ and\ \bibinfo {author} {\bibfnamefont {E.~V.}\
			\bibnamefont {Frolova}},\ }\href@noop {} {\bibfield  {journal} {\bibinfo
			{journal} {Phys. Rev. B}\ }\textbf {\bibinfo {volume} {78}},\ \bibinfo
		{pages} {235321} (\bibinfo {year} {2008})}\BibitemShut {NoStop}%
	\bibitem [{\citenamefont {Rusin}\ and\ \citenamefont
		{Zawadzki}(2009)}]{Rusin09}%
	\BibitemOpen
	\bibfield  {author} {\bibinfo {author} {\bibfnamefont {T.~M.}\ \bibnamefont
			{Rusin}}\ and\ \bibinfo {author} {\bibfnamefont {W.}~\bibnamefont
			{Zawadzki}},\ }\href@noop {} {\bibfield  {journal} {\bibinfo  {journal}
			{Phys. Rev. B}\ }\textbf {\bibinfo {volume} {80}},\ \bibinfo {pages} {045416}
		(\bibinfo {year} {2009})}\BibitemShut {NoStop}%
	\bibitem [{\citenamefont {Wang}\ \emph {et~al.}(2010)\citenamefont {Wang},
		\citenamefont {Yang},\ and\ \citenamefont {Xiong}}]{Wang10}%
	\BibitemOpen
	\bibfield  {author} {\bibinfo {author} {\bibfnamefont {Y.-X.}\ \bibnamefont
			{Wang}}, \bibinfo {author} {\bibfnamefont {Z.}~\bibnamefont {Yang}}, \ and\
		\bibinfo {author} {\bibfnamefont {S.-J.}\ \bibnamefont {Xiong}},\ }\href@noop
	{} {\bibfield  {journal} {\bibinfo  {journal} {EPL (Europhysics Letters)}\
		}\textbf {\bibinfo {volume} {89}},\ \bibinfo {pages} {17007} (\bibinfo {year}
		{2010})}\BibitemShut {NoStop}%
	\bibitem [{\citenamefont {Martinez}\ \emph {et~al.}(2010)\citenamefont
		{Martinez}, \citenamefont {Jalil},\ and\ \citenamefont
		{Tan}}]{martinez2010klein}%
	\BibitemOpen
	\bibfield  {author} {\bibinfo {author} {\bibfnamefont {J.}~\bibnamefont
			{Martinez}}, \bibinfo {author} {\bibfnamefont {M.}~\bibnamefont {Jalil}}, \
		and\ \bibinfo {author} {\bibfnamefont {S.}~\bibnamefont {Tan}},\ }\href@noop
	{} {\bibfield  {journal} {\bibinfo  {journal} {Applied Physics Letters}\
		}\textbf {\bibinfo {volume} {97}},\ \bibinfo {pages} {062111} (\bibinfo
		{year} {2010})}\BibitemShut {NoStop}%
	\bibitem [{\citenamefont {Rusin}\ and\ \citenamefont
		{Zawadzki}(2014)}]{Rusin14}%
	\BibitemOpen
	\bibfield  {author} {\bibinfo {author} {\bibfnamefont {T.~M.}\ \bibnamefont
			{Rusin}}\ and\ \bibinfo {author} {\bibfnamefont {W.}~\bibnamefont
			{Zawadzki}},\ }\href@noop {} {\bibfield  {journal} {\bibinfo  {journal}
			{Journal of Physics: Condensed Matter}\ }\textbf {\bibinfo {volume} {26}},\
		\bibinfo {pages} {215301} (\bibinfo {year} {2014})}\BibitemShut {NoStop}%
	\bibitem [{\citenamefont {Yanyushkina}\ \emph {et~al.}(2012)\citenamefont
		{Yanyushkina}, \citenamefont {Zhukov}, \citenamefont {Belonenko},\ and\
		\citenamefont {George}}]{yanyushkina2012zitterbewegung}%
	\BibitemOpen
	\bibfield  {author} {\bibinfo {author} {\bibfnamefont {N.~N.}\ \bibnamefont
			{Yanyushkina}}, \bibinfo {author} {\bibfnamefont {A.~V.}\ \bibnamefont
			{Zhukov}}, \bibinfo {author} {\bibfnamefont {M.~B.}\ \bibnamefont
			{Belonenko}}, \ and\ \bibinfo {author} {\bibfnamefont {T.~F.}\ \bibnamefont
			{George}},\ }\href@noop {} {\bibfield  {journal} {\bibinfo  {journal} {Modern
				Physics Letters B}\ }\textbf {\bibinfo {volume} {26}},\ \bibinfo {pages}
		{1250106} (\bibinfo {year} {2012})}\BibitemShut {NoStop}%
	\bibitem [{\citenamefont {Ferreira}\ \emph {et~al.}(2018)\citenamefont
		{Ferreira}, \citenamefont {Maciel}, \citenamefont {Penteado},\ and\
		\citenamefont {Egues}}]{PhysRevB.98.165120}%
	\BibitemOpen
	\bibfield  {author} {\bibinfo {author} {\bibfnamefont {G.~J.}\ \bibnamefont
			{Ferreira}}, \bibinfo {author} {\bibfnamefont {R.~P.}\ \bibnamefont
			{Maciel}}, \bibinfo {author} {\bibfnamefont {P.~H.}\ \bibnamefont
			{Penteado}}, \ and\ \bibinfo {author} {\bibfnamefont {J.~C.}\ \bibnamefont
			{Egues}},\ }\href {\doibase 10.1103/PhysRevB.98.165120} {\bibfield  {journal}
		{\bibinfo  {journal} {Phys. Rev. B}\ }\textbf {\bibinfo {volume} {98}},\
		\bibinfo {pages} {165120} (\bibinfo {year} {2018})}\BibitemShut {NoStop}%
	\bibitem [{\citenamefont {Huang}\ \emph {et~al.}(2018)\citenamefont {Huang},
		\citenamefont {Ma},\ and\ \citenamefont {Wang}}]{huang2018zitterbewegung}%
	\BibitemOpen
	\bibfield  {author} {\bibinfo {author} {\bibfnamefont {T.}~\bibnamefont
			{Huang}}, \bibinfo {author} {\bibfnamefont {T.}~\bibnamefont {Ma}}, \ and\
		\bibinfo {author} {\bibfnamefont {L.-G.}\ \bibnamefont {Wang}},\ }\href@noop
	{} {\bibfield  {journal} {\bibinfo  {journal} {Journal of Physics: Condensed
				Matter}\ }\textbf {\bibinfo {volume} {30}},\ \bibinfo {pages} {245501}
		(\bibinfo {year} {2018})}\BibitemShut {NoStop}%
	\bibitem [{\citenamefont {Cannata}\ \emph {et~al.}(1990)\citenamefont
		{Cannata}, \citenamefont {Ferrari},\ and\ \citenamefont {Russo}}]{Cannata90}%
	\BibitemOpen
	\bibfield  {author} {\bibinfo {author} {\bibfnamefont {F.}~\bibnamefont
			{Cannata}}, \bibinfo {author} {\bibfnamefont {L.}~\bibnamefont {Ferrari}}, \
		and\ \bibinfo {author} {\bibfnamefont {G.}~\bibnamefont {Russo}},\
	}\href@noop {} {\bibfield  {journal} {\bibinfo  {journal} {Solid State
				Commun.}\ }\textbf {\bibinfo {volume} {74}},\ \bibinfo {pages} {309}
		(\bibinfo {year} {1990})}\BibitemShut {NoStop}%
	\bibitem [{\citenamefont {Lurie}\ and\ \citenamefont {Cremer}(1970)}]{Lurie70}%
	\BibitemOpen
	\bibfield  {author} {\bibinfo {author} {\bibfnamefont {D.}~\bibnamefont
			{Lurie}}\ and\ \bibinfo {author} {\bibfnamefont {S.}~\bibnamefont {Cremer}},\
	}\href@noop {} {\bibfield  {journal} {\bibinfo  {journal} {Physica
				(Amsterdam)}\ }\textbf {\bibinfo {volume} {50}},\ \bibinfo {pages} {224}
		(\bibinfo {year} {1970})}\BibitemShut {NoStop}%
	\bibitem [{\citenamefont {LeBlanc}\ \emph {et~al.}(2013)\citenamefont
		{LeBlanc}, \citenamefont {Beeler}, \citenamefont {Jimenez-Garcia},
		\citenamefont {Perry}, \citenamefont {Sugawa}, \citenamefont {Williams},\
		and\ \citenamefont {Spielman}}]{leblanc2013direct}%
	\BibitemOpen
	\bibfield  {author} {\bibinfo {author} {\bibfnamefont {L.~J.}\ \bibnamefont
			{LeBlanc}}, \bibinfo {author} {\bibfnamefont {M.}~\bibnamefont {Beeler}},
		\bibinfo {author} {\bibfnamefont {K.}~\bibnamefont {Jimenez-Garcia}},
		\bibinfo {author} {\bibfnamefont {A.~R.}\ \bibnamefont {Perry}}, \bibinfo
		{author} {\bibfnamefont {S.}~\bibnamefont {Sugawa}}, \bibinfo {author}
		{\bibfnamefont {R.}~\bibnamefont {Williams}}, \ and\ \bibinfo {author}
		{\bibfnamefont {I.~B.}\ \bibnamefont {Spielman}},\ }\href@noop {} {\bibfield
		{journal} {\bibinfo  {journal} {New Journal of Physics}\ }\textbf {\bibinfo
			{volume} {15}},\ \bibinfo {pages} {073011} (\bibinfo {year}
		{2013})}\BibitemShut {NoStop}%
	\bibitem [{\citenamefont {Zhang}\ \emph {et~al.}(2013)\citenamefont {Zhang},
		\citenamefont {Song}, \citenamefont {Liu},\ and\ \citenamefont
		{Liu}}]{Zhang13}%
	\BibitemOpen
	\bibfield  {author} {\bibinfo {author} {\bibfnamefont {Y.-C.}\ \bibnamefont
			{Zhang}}, \bibinfo {author} {\bibfnamefont {S.-W.}\ \bibnamefont {Song}},
		\bibinfo {author} {\bibfnamefont {C.-F.}\ \bibnamefont {Liu}}, \ and\
		\bibinfo {author} {\bibfnamefont {W.-M.}\ \bibnamefont {Liu}},\ }\href@noop
	{} {\bibfield  {journal} {\bibinfo  {journal} {Phys. Rev. A}\ }\textbf
		{\bibinfo {volume} {87}},\ \bibinfo {pages} {023612} (\bibinfo {year}
		{2013})}\BibitemShut {NoStop}%
	\bibitem [{\citenamefont {Qu}\ \emph {et~al.}(2013{\natexlab{a}})\citenamefont
		{Qu}, \citenamefont {Hamner}, \citenamefont {Gong}, \citenamefont {Zhang},\
		and\ \citenamefont {Engels}}]{PhysRevA.88.021604}%
	\BibitemOpen
	\bibfield  {author} {\bibinfo {author} {\bibfnamefont {C.}~\bibnamefont
			{Qu}}, \bibinfo {author} {\bibfnamefont {C.}~\bibnamefont {Hamner}}, \bibinfo
		{author} {\bibfnamefont {M.}~\bibnamefont {Gong}}, \bibinfo {author}
		{\bibfnamefont {C.}~\bibnamefont {Zhang}}, \ and\ \bibinfo {author}
		{\bibfnamefont {P.}~\bibnamefont {Engels}},\ }\href {\doibase
		10.1103/PhysRevA.88.021604} {\bibfield  {journal} {\bibinfo  {journal} {Phys.
				Rev. A}\ }\textbf {\bibinfo {volume} {88}},\ \bibinfo {pages} {021604}
		(\bibinfo {year} {2013}{\natexlab{a}})}\BibitemShut {NoStop}%
	\bibitem [{\citenamefont {Vaishnav}\ and\ \citenamefont
		{Clark}(2008)}]{Vaish08}%
	\BibitemOpen
	\bibfield  {author} {\bibinfo {author} {\bibfnamefont {J.~Y.}\ \bibnamefont
			{Vaishnav}}\ and\ \bibinfo {author} {\bibfnamefont {C.~W.}\ \bibnamefont
			{Clark}},\ }\href@noop {} {\bibfield  {journal} {\bibinfo  {journal} {Phys.
				Rev. Lett.}\ }\textbf {\bibinfo {volume} {100}},\ \bibinfo {pages} {153002}
		(\bibinfo {year} {2008})}\BibitemShut {NoStop}%
	\bibitem [{\citenamefont {Lamata}\ \emph {et~al.}(2007)\citenamefont {Lamata},
		\citenamefont {Le\'on}, \citenamefont {Sch\"atz},\ and\ \citenamefont
		{Solano}}]{Lamata07}%
	\BibitemOpen
	\bibfield  {author} {\bibinfo {author} {\bibfnamefont {L.}~\bibnamefont
			{Lamata}}, \bibinfo {author} {\bibfnamefont {J.}~\bibnamefont {Le\'on}},
		\bibinfo {author} {\bibfnamefont {T.}~\bibnamefont {Sch\"atz}}, \ and\
		\bibinfo {author} {\bibfnamefont {E.}~\bibnamefont {Solano}},\ }\href@noop {}
	{\bibfield  {journal} {\bibinfo  {journal} {Phys. Rev. Lett.}\ }\textbf
		{\bibinfo {volume} {98}},\ \bibinfo {pages} {253005} (\bibinfo {year}
		{2007})}\BibitemShut {NoStop}%
	\bibitem [{\citenamefont {Bermudez}\ \emph {et~al.}(2007)\citenamefont
		{Bermudez}, \citenamefont {Martin-Delgado},\ and\ \citenamefont
		{Solano}}]{Bermu07}%
	\BibitemOpen
	\bibfield  {author} {\bibinfo {author} {\bibfnamefont {A.}~\bibnamefont
			{Bermudez}}, \bibinfo {author} {\bibfnamefont {M.~A.}\ \bibnamefont
			{Martin-Delgado}}, \ and\ \bibinfo {author} {\bibfnamefont {E.}~\bibnamefont
			{Solano}},\ }\href@noop {} {\bibfield  {journal} {\bibinfo  {journal} {Phys.
				Rev. A}\ }\textbf {\bibinfo {volume} {76}},\ \bibinfo {pages} {041801}
		(\bibinfo {year} {2007})}\BibitemShut {NoStop}%
	\bibitem [{\citenamefont {Gerritsma}\ \emph {et~al.}(2007)\citenamefont
		{Gerritsma}, \citenamefont {Kirchmair}, \citenamefont {Zahringer},
		\citenamefont {E.~Solano},\ and\ \citenamefont {Roos}}]{Gerrit10}%
	\BibitemOpen
	\bibfield  {author} {\bibinfo {author} {\bibfnamefont {R.}~\bibnamefont
			{Gerritsma}}, \bibinfo {author} {\bibfnamefont {G.}~\bibnamefont
			{Kirchmair}}, \bibinfo {author} {\bibfnamefont {F.}~\bibnamefont
			{Zahringer}}, \bibinfo {author} {\bibfnamefont {R.~B.}\ \bibnamefont
			{E.~Solano}}, \ and\ \bibinfo {author} {\bibfnamefont {C.~F.}\ \bibnamefont
			{Roos}},\ }\href@noop {} {\bibfield  {journal} {\bibinfo  {journal} {Nature
				(London)}\ }\textbf {\bibinfo {volume} {463}},\ \bibinfo {pages} {68}
		(\bibinfo {year} {2007})}\BibitemShut {NoStop}%
	\bibitem [{\citenamefont {Qu}\ \emph {et~al.}(2013{\natexlab{b}})\citenamefont
		{Qu}, \citenamefont {Hamner}, \citenamefont {Gong}, \citenamefont {Zhang},\
		and\ \citenamefont {Engels}}]{Qu13}%
	\BibitemOpen
	\bibfield  {author} {\bibinfo {author} {\bibfnamefont {C.}~\bibnamefont
			{Qu}}, \bibinfo {author} {\bibfnamefont {C.}~\bibnamefont {Hamner}}, \bibinfo
		{author} {\bibfnamefont {M.}~\bibnamefont {Gong}}, \bibinfo {author}
		{\bibfnamefont {C.}~\bibnamefont {Zhang}}, \ and\ \bibinfo {author}
		{\bibfnamefont {P.}~\bibnamefont {Engels}},\ }\href@noop {} {\bibfield
		{journal} {\bibinfo  {journal} {Phys. Rev. A}\ }\textbf {\bibinfo {volume}
			{88}},\ \bibinfo {pages} {021604} (\bibinfo {year}
		{2013}{\natexlab{b}})}\BibitemShut {NoStop}%
	\bibitem [{\citenamefont {Stepanov}\ \emph {et~al.}(2016)\citenamefont
		{Stepanov}, \citenamefont {Ersfeld}, \citenamefont {Poshakinskiy},
		\citenamefont {Lepsa}, \citenamefont {Ivchenko}, \citenamefont {Tarasenko},\
		and\ \citenamefont {Beschoten}}]{stepanov2016coherent}%
	\BibitemOpen
	\bibfield  {author} {\bibinfo {author} {\bibfnamefont {I.}~\bibnamefont
			{Stepanov}}, \bibinfo {author} {\bibfnamefont {M.}~\bibnamefont {Ersfeld}},
		\bibinfo {author} {\bibfnamefont {A.}~\bibnamefont {Poshakinskiy}}, \bibinfo
		{author} {\bibfnamefont {M.}~\bibnamefont {Lepsa}}, \bibinfo {author}
		{\bibfnamefont {E.}~\bibnamefont {Ivchenko}}, \bibinfo {author}
		{\bibfnamefont {S.}~\bibnamefont {Tarasenko}}, \ and\ \bibinfo {author}
		{\bibfnamefont {B.}~\bibnamefont {Beschoten}},\ }\href@noop {} {\bibfield
		{journal} {\bibinfo  {journal} {arXiv preprint arXiv:1612.06190}\ } (\bibinfo
		{year} {2016})}\BibitemShut {NoStop}%
	\bibitem [{\citenamefont {Iwasaki}\ \emph {et~al.}(2017)\citenamefont
		{Iwasaki}, \citenamefont {Hashimoto}, \citenamefont {Nakamura},\ and\
		\citenamefont {Katsumoto}}]{iwasaki2017observation}%
	\BibitemOpen
	\bibfield  {author} {\bibinfo {author} {\bibfnamefont {Y.}~\bibnamefont
			{Iwasaki}}, \bibinfo {author} {\bibfnamefont {Y.}~\bibnamefont {Hashimoto}},
		\bibinfo {author} {\bibfnamefont {T.}~\bibnamefont {Nakamura}}, \ and\
		\bibinfo {author} {\bibfnamefont {S.}~\bibnamefont {Katsumoto}},\ }in\
	\href@noop {} {\emph {\bibinfo {booktitle} {Journal of Physics: Conference
				Series}}},\ Vol.\ \bibinfo {volume} {864}\ (\bibinfo {organization} {IOP
		Publishing},\ \bibinfo {year} {2017})\ p.\ \bibinfo {pages}
	{012054}\BibitemShut {NoStop}%
	\bibitem [{\citenamefont {Huang}(1952)}]{huang1952zitterbewegung}%
	\BibitemOpen
	\bibfield  {author} {\bibinfo {author} {\bibfnamefont {K.}~\bibnamefont
			{Huang}},\ }\href@noop {} {\bibfield  {journal} {\bibinfo  {journal}
			{American Journal of Physics}\ }\textbf {\bibinfo {volume} {20}},\ \bibinfo
		{pages} {479} (\bibinfo {year} {1952})}\BibitemShut {NoStop}%
	\bibitem [{\citenamefont {Lock}(1979)}]{lock1979zitterbewegung}%
	\BibitemOpen
	\bibfield  {author} {\bibinfo {author} {\bibfnamefont {J.~A.}\ \bibnamefont
			{Lock}},\ }\href@noop {} {\bibfield  {journal} {\bibinfo  {journal} {American
				Journal of Physics}\ }\textbf {\bibinfo {volume} {47}},\ \bibinfo {pages}
		{797} (\bibinfo {year} {1979})}\BibitemShut {NoStop}%
	\bibitem [{\citenamefont {Zawadzki}(2005)}]{PhysRevB.72.085217}%
	\BibitemOpen
	\bibfield  {author} {\bibinfo {author} {\bibfnamefont {W.}~\bibnamefont
			{Zawadzki}},\ }\href {\doibase 10.1103/PhysRevB.72.085217} {\bibfield
		{journal} {\bibinfo  {journal} {Phys. Rev. B}\ }\textbf {\bibinfo {volume}
			{72}},\ \bibinfo {pages} {085217} (\bibinfo {year} {2005})}\BibitemShut
	{NoStop}%
	\bibitem [{\citenamefont {Ho}\ \emph {et~al.}(2014)\citenamefont {Ho},
		\citenamefont {Jalil},\ and\ \citenamefont {Tan}}]{ho2014persistent}%
	\BibitemOpen
	\bibfield  {author} {\bibinfo {author} {\bibfnamefont {C.~S.}\ \bibnamefont
			{Ho}}, \bibinfo {author} {\bibfnamefont {M.~B.}\ \bibnamefont {Jalil}}, \
		and\ \bibinfo {author} {\bibfnamefont {S.~G.}\ \bibnamefont {Tan}},\
	}\href@noop {} {\bibfield  {journal} {\bibinfo  {journal} {EPL (Europhysics
				Letters)}\ }\textbf {\bibinfo {volume} {108}},\ \bibinfo {pages} {27012}
		(\bibinfo {year} {2014})}\BibitemShut {NoStop}%
	\bibitem [{\citenamefont {Reck}\ \emph {et~al.}(2020)\citenamefont {Reck},
		\citenamefont {Gorini},\ and\ \citenamefont {Richter}}]{PhysRevB.101.094306}%
	\BibitemOpen
	\bibfield  {author} {\bibinfo {author} {\bibfnamefont {P.}~\bibnamefont
			{Reck}}, \bibinfo {author} {\bibfnamefont {C.}~\bibnamefont {Gorini}}, \ and\
		\bibinfo {author} {\bibfnamefont {K.}~\bibnamefont {Richter}},\ }\href
	{\doibase 10.1103/PhysRevB.101.094306} {\bibfield  {journal} {\bibinfo
			{journal} {Phys. Rev. B}\ }\textbf {\bibinfo {volume} {101}},\ \bibinfo
		{pages} {094306} (\bibinfo {year} {2020})}\BibitemShut {NoStop}%
	\bibitem [{\citenamefont {Zyuzin}\ \emph {et~al.}(2011)\citenamefont {Zyuzin},
		\citenamefont {Hook},\ and\ \citenamefont {Burkov}}]{zyuzin2011parallel}%
	\BibitemOpen
	\bibfield  {author} {\bibinfo {author} {\bibfnamefont {A.}~\bibnamefont
			{Zyuzin}}, \bibinfo {author} {\bibfnamefont {M.}~\bibnamefont {Hook}}, \ and\
		\bibinfo {author} {\bibfnamefont {A.}~\bibnamefont {Burkov}},\ }\href@noop {}
	{\bibfield  {journal} {\bibinfo  {journal} {Physical Review B}\ }\textbf
		{\bibinfo {volume} {83}},\ \bibinfo {pages} {245428} (\bibinfo {year}
		{2011})}\BibitemShut {NoStop}%
	\bibitem [{\citenamefont {Linder}\ \emph {et~al.}(2009)\citenamefont {Linder},
		\citenamefont {Yokoyama},\ and\ \citenamefont {Sudb\o{}}}]{Linder:prb09}%
	\BibitemOpen
	\bibfield  {author} {\bibinfo {author} {\bibfnamefont {J.}~\bibnamefont
			{Linder}}, \bibinfo {author} {\bibfnamefont {T.}~\bibnamefont {Yokoyama}}, \
		and\ \bibinfo {author} {\bibfnamefont {A.}~\bibnamefont {Sudb\o{}}},\ }\href
	{\doibase 10.1103/PhysRevB.80.205401} {\bibfield  {journal} {\bibinfo
			{journal} {Phys. Rev. B}\ }\textbf {\bibinfo {volume} {80}},\ \bibinfo
		{pages} {205401} (\bibinfo {year} {2009})}\BibitemShut {NoStop}%
	\bibitem [{\citenamefont {Lu}\ \emph {et~al.}(2010)\citenamefont {Lu},
		\citenamefont {Shan}, \citenamefont {Yao}, \citenamefont {Niu},\ and\
		\citenamefont {Shen}}]{Lu:prb10}%
	\BibitemOpen
	\bibfield  {author} {\bibinfo {author} {\bibfnamefont {H.-Z.}\ \bibnamefont
			{Lu}}, \bibinfo {author} {\bibfnamefont {W.-Y.}\ \bibnamefont {Shan}},
		\bibinfo {author} {\bibfnamefont {W.}~\bibnamefont {Yao}}, \bibinfo {author}
		{\bibfnamefont {Q.}~\bibnamefont {Niu}}, \ and\ \bibinfo {author}
		{\bibfnamefont {S.-Q.}\ \bibnamefont {Shen}},\ }\href
	{http://link.aps.org/doi/10.1103/PhysRevB.81.115407} {\bibfield  {journal}
		{\bibinfo  {journal} {Phys. Rev. B}\ }\textbf {\bibinfo {volume} {81}},\
		\bibinfo {pages} {115407} (\bibinfo {year} {2010})}\BibitemShut {NoStop}%
	\bibitem [{\citenamefont {Liu}\ \emph {et~al.}(2010)\citenamefont {Liu},
		\citenamefont {Zhang}, \citenamefont {Yan}, \citenamefont {Qi}, \citenamefont
		{Frauenheim}, \citenamefont {Dai}, \citenamefont {Fang},\ and\ \citenamefont
		{Zhang}}]{Liu:prb10}%
	\BibitemOpen
	\bibfield  {author} {\bibinfo {author} {\bibfnamefont {C.-X.}\ \bibnamefont
			{Liu}}, \bibinfo {author} {\bibfnamefont {H.}~\bibnamefont {Zhang}}, \bibinfo
		{author} {\bibfnamefont {B.}~\bibnamefont {Yan}}, \bibinfo {author}
		{\bibfnamefont {X.-L.}\ \bibnamefont {Qi}}, \bibinfo {author} {\bibfnamefont
			{T.}~\bibnamefont {Frauenheim}}, \bibinfo {author} {\bibfnamefont
			{X.}~\bibnamefont {Dai}}, \bibinfo {author} {\bibfnamefont {Z.}~\bibnamefont
			{Fang}}, \ and\ \bibinfo {author} {\bibfnamefont {S.-C.}\ \bibnamefont
			{Zhang}},\ }\href {http://link.aps.org/doi/10.1103/PhysRevB.81.041307}
	{\bibfield  {journal} {\bibinfo  {journal} {Phys. Rev. B}\ }\textbf {\bibinfo
			{volume} {81}},\ \bibinfo {pages} {041307} (\bibinfo {year}
		{2010})}\BibitemShut {NoStop}%
	\bibitem [{\citenamefont {Ruderman}\ and\ \citenamefont
		{Kittel}(1954)}]{RKKY1}%
	\BibitemOpen
	\bibfield  {author} {\bibinfo {author} {\bibfnamefont {M.~A.}\ \bibnamefont
			{Ruderman}}\ and\ \bibinfo {author} {\bibfnamefont {C.}~\bibnamefont
			{Kittel}},\ }\href {\doibase 10.1103/PhysRev.96.99} {\bibfield  {journal}
		{\bibinfo  {journal} {Phys. Rev.}\ }\textbf {\bibinfo {volume} {96}},\
		\bibinfo {pages} {99} (\bibinfo {year} {1954})}\BibitemShut {NoStop}%
	\bibitem [{\citenamefont {Kasuya}(1956)}]{RKKY2}%
	\BibitemOpen
	\bibfield  {author} {\bibinfo {author} {\bibfnamefont {T.}~\bibnamefont
			{Kasuya}},\ }\href@noop {} {\bibfield  {journal} {\bibinfo  {journal}
			{Progress of theoretical physics}\ }\textbf {\bibinfo {volume} {16}},\
		\bibinfo {pages} {45} (\bibinfo {year} {1956})}\BibitemShut {NoStop}%
	\bibitem [{\citenamefont {Yosida}(1957)}]{RKKY3}%
	\BibitemOpen
	\bibfield  {author} {\bibinfo {author} {\bibfnamefont {K.}~\bibnamefont
			{Yosida}},\ }\href {\doibase 10.1103/PhysRev.106.893} {\bibfield  {journal}
		{\bibinfo  {journal} {Phys. Rev.}\ }\textbf {\bibinfo {volume} {106}},\
		\bibinfo {pages} {893} (\bibinfo {year} {1957})}\BibitemShut {NoStop}%
	\bibitem [{\citenamefont {Mattis}(2006)}]{mattis2006theory}%
	\BibitemOpen
	\bibfield  {author} {\bibinfo {author} {\bibfnamefont {D.~C.}\ \bibnamefont
			{Mattis}},\ }\href@noop {} {\emph {\bibinfo {title} {The theory of magnetism
				made simple}}}\ (\bibinfo  {publisher} {World Scientific Publishing
		Company},\ \bibinfo {year} {2006})\BibitemShut {NoStop}%
	\bibitem [{\citenamefont {Tagirov}(1993)}]{TAGIROV1993257}%
	\BibitemOpen
	\bibfield  {author} {\bibinfo {author} {\bibfnamefont {L.}~\bibnamefont
			{Tagirov}},\ }\href {\doibase https://doi.org/10.1016/0038-1098(93)90754-B}
	{\bibfield  {journal} {\bibinfo  {journal} {Solid State Communications}\
		}\textbf {\bibinfo {volume} {88}},\ \bibinfo {pages} {257 } (\bibinfo {year}
		{1993})}\BibitemShut {NoStop}%
	\bibitem [{\citenamefont {Akbari}\ \emph {et~al.}(2011)\citenamefont {Akbari},
		\citenamefont {Eremin},\ and\ \citenamefont {Thalmeier}}]{akbari2011rkky}%
	\BibitemOpen
	\bibfield  {author} {\bibinfo {author} {\bibfnamefont {A.}~\bibnamefont
			{Akbari}}, \bibinfo {author} {\bibfnamefont {I.}~\bibnamefont {Eremin}}, \
		and\ \bibinfo {author} {\bibfnamefont {P.}~\bibnamefont {Thalmeier}},\
	}\href@noop {} {\bibfield  {journal} {\bibinfo  {journal} {Physical Review
				B}\ }\textbf {\bibinfo {volume} {84}},\ \bibinfo {pages} {134513} (\bibinfo
		{year} {2011})}\BibitemShut {NoStop}%
	\bibitem [{\citenamefont {Aristov}\ \emph {et~al.}(1997)\citenamefont
		{Aristov}, \citenamefont {Maleyev},\ and\ \citenamefont
		{Yashenkin}}]{aristov1997rkky}%
	\BibitemOpen
	\bibfield  {author} {\bibinfo {author} {\bibfnamefont {D.}~\bibnamefont
			{Aristov}}, \bibinfo {author} {\bibfnamefont {S.}~\bibnamefont {Maleyev}}, \
		and\ \bibinfo {author} {\bibfnamefont {A.}~\bibnamefont {Yashenkin}},\
	}\href@noop {} {\bibfield  {journal} {\bibinfo  {journal} {Zeitschrift
				f{\"u}r Physik B Condensed Matter}\ }\textbf {\bibinfo {volume} {102}},\
		\bibinfo {pages} {467} (\bibinfo {year} {1997})}\BibitemShut {NoStop}%
	\bibitem [{\citenamefont {Zhu}\ \emph {et~al.}(2011{\natexlab{a}})\citenamefont
		{Zhu}, \citenamefont {Yao}, \citenamefont {Zhang},\ and\ \citenamefont
		{Chang}}]{RKKYTI:PRL2011}%
	\BibitemOpen
	\bibfield  {author} {\bibinfo {author} {\bibfnamefont {J.-J.}\ \bibnamefont
			{Zhu}}, \bibinfo {author} {\bibfnamefont {D.-X.}\ \bibnamefont {Yao}},
		\bibinfo {author} {\bibfnamefont {S.-C.}\ \bibnamefont {Zhang}}, \ and\
		\bibinfo {author} {\bibfnamefont {K.}~\bibnamefont {Chang}},\ }\href
	{\doibase 10.1103/PhysRevLett.106.097201} {\bibfield  {journal} {\bibinfo
			{journal} {Phys. Rev. Lett.}\ }\textbf {\bibinfo {volume} {106}},\ \bibinfo
		{pages} {097201} (\bibinfo {year} {2011}{\natexlab{a}})}\BibitemShut
	{NoStop}%
	\bibitem [{\citenamefont {Biswas}\ and\ \citenamefont
		{Balatsky}(2010)}]{PhysRevB.81.233405}%
	\BibitemOpen
	\bibfield  {author} {\bibinfo {author} {\bibfnamefont {R.~R.}\ \bibnamefont
			{Biswas}}\ and\ \bibinfo {author} {\bibfnamefont {A.~V.}\ \bibnamefont
			{Balatsky}},\ }\href {\doibase 10.1103/PhysRevB.81.233405} {\bibfield
		{journal} {\bibinfo  {journal} {Phys. Rev. B}\ }\textbf {\bibinfo {volume}
			{81}},\ \bibinfo {pages} {233405} (\bibinfo {year} {2010})}\BibitemShut
	{NoStop}%
	\bibitem [{\citenamefont {Garate}\ and\ \citenamefont
		{Franz}(2010)}]{PhysRevB.81.172408}%
	\BibitemOpen
	\bibfield  {author} {\bibinfo {author} {\bibfnamefont {I.}~\bibnamefont
			{Garate}}\ and\ \bibinfo {author} {\bibfnamefont {M.}~\bibnamefont {Franz}},\
	}\href {\doibase 10.1103/PhysRevB.81.172408} {\bibfield  {journal} {\bibinfo
			{journal} {Phys. Rev. B}\ }\textbf {\bibinfo {volume} {81}},\ \bibinfo
		{pages} {172408} (\bibinfo {year} {2010})}\BibitemShut {NoStop}%
	\bibitem [{\citenamefont {Zhu}\ \emph {et~al.}(2011{\natexlab{b}})\citenamefont
		{Zhu}, \citenamefont {Yao}, \citenamefont {Zhang},\ and\ \citenamefont
		{Chang}}]{PhysRevLett.106.097201}%
	\BibitemOpen
	\bibfield  {author} {\bibinfo {author} {\bibfnamefont {J.-J.}\ \bibnamefont
			{Zhu}}, \bibinfo {author} {\bibfnamefont {D.-X.}\ \bibnamefont {Yao}},
		\bibinfo {author} {\bibfnamefont {S.-C.}\ \bibnamefont {Zhang}}, \ and\
		\bibinfo {author} {\bibfnamefont {K.}~\bibnamefont {Chang}},\ }\href
	{\doibase 10.1103/PhysRevLett.106.097201} {\bibfield  {journal} {\bibinfo
			{journal} {Phys. Rev. Lett.}\ }\textbf {\bibinfo {volume} {106}},\ \bibinfo
		{pages} {097201} (\bibinfo {year} {2011}{\natexlab{b}})}\BibitemShut
	{NoStop}%
	\bibitem [{\citenamefont {Abanin}\ and\ \citenamefont
		{Pesin}(2011)}]{PhysRevLett.106.136802}%
	\BibitemOpen
	\bibfield  {author} {\bibinfo {author} {\bibfnamefont {D.~A.}\ \bibnamefont
			{Abanin}}\ and\ \bibinfo {author} {\bibfnamefont {D.~A.}\ \bibnamefont
			{Pesin}},\ }\href {\doibase 10.1103/PhysRevLett.106.136802} {\bibfield
		{journal} {\bibinfo  {journal} {Phys. Rev. Lett.}\ }\textbf {\bibinfo
			{volume} {106}},\ \bibinfo {pages} {136802} (\bibinfo {year}
		{2011})}\BibitemShut {NoStop}%
	\bibitem [{\citenamefont {Zyuzin}\ and\ \citenamefont
		{Loss}(2014)}]{PhysRevB.90.125443}%
	\BibitemOpen
	\bibfield  {author} {\bibinfo {author} {\bibfnamefont {A.~A.}\ \bibnamefont
			{Zyuzin}}\ and\ \bibinfo {author} {\bibfnamefont {D.}~\bibnamefont {Loss}},\
	}\href {\doibase 10.1103/PhysRevB.90.125443} {\bibfield  {journal} {\bibinfo
			{journal} {Phys. Rev. B}\ }\textbf {\bibinfo {volume} {90}},\ \bibinfo
		{pages} {125443} (\bibinfo {year} {2014})}\BibitemShut {NoStop}%
	\bibitem [{\citenamefont {Chang}\ \emph {et~al.}(2015)\citenamefont {Chang},
		\citenamefont {Zhou}, \citenamefont {Wang}, \citenamefont {Shan},\ and\
		\citenamefont {Xiao}}]{PhysRevB.92.241103}%
	\BibitemOpen
	\bibfield  {author} {\bibinfo {author} {\bibfnamefont {H.-R.}\ \bibnamefont
			{Chang}}, \bibinfo {author} {\bibfnamefont {J.}~\bibnamefont {Zhou}},
		\bibinfo {author} {\bibfnamefont {S.-X.}\ \bibnamefont {Wang}}, \bibinfo
		{author} {\bibfnamefont {W.-Y.}\ \bibnamefont {Shan}}, \ and\ \bibinfo
		{author} {\bibfnamefont {D.}~\bibnamefont {Xiao}},\ }\href {\doibase
		10.1103/PhysRevB.92.241103} {\bibfield  {journal} {\bibinfo  {journal} {Phys.
				Rev. B}\ }\textbf {\bibinfo {volume} {92}},\ \bibinfo {pages} {241103}
		(\bibinfo {year} {2015})}\BibitemShut {NoStop}%
	\bibitem [{\citenamefont {Sun}\ and\ \citenamefont {Wang}(2017)}]{Sun_2017}%
	\BibitemOpen
	\bibfield  {author} {\bibinfo {author} {\bibfnamefont {Y.}~\bibnamefont
			{Sun}}\ and\ \bibinfo {author} {\bibfnamefont {A.}~\bibnamefont {Wang}},\
	}\href {\doibase 10.1088/1361-648x/aa8932} {\bibfield  {journal} {\bibinfo
			{journal} {Journal of Physics: Condensed Matter}\ }\textbf {\bibinfo {volume}
			{29}},\ \bibinfo {pages} {435306} (\bibinfo {year} {2017})}\BibitemShut
	{NoStop}%
	\bibitem [{\citenamefont {Verma}\ \emph {et~al.}(2020)\citenamefont {Verma},
		\citenamefont {Giri}, \citenamefont {Fertig},\ and\ \citenamefont
		{Kundu}}]{PhysRevB.101.085419}%
	\BibitemOpen
	\bibfield  {author} {\bibinfo {author} {\bibfnamefont {S.}~\bibnamefont
			{Verma}}, \bibinfo {author} {\bibfnamefont {D.}~\bibnamefont {Giri}},
		\bibinfo {author} {\bibfnamefont {H.~A.}\ \bibnamefont {Fertig}}, \ and\
		\bibinfo {author} {\bibfnamefont {A.}~\bibnamefont {Kundu}},\ }\href
	{\doibase 10.1103/PhysRevB.101.085419} {\bibfield  {journal} {\bibinfo
			{journal} {Phys. Rev. B}\ }\textbf {\bibinfo {volume} {101}},\ \bibinfo
		{pages} {085419} (\bibinfo {year} {2020})}\BibitemShut {NoStop}%
	\bibitem [{\citenamefont {Kaladzhyan}\ \emph {et~al.}(2019)\citenamefont
		{Kaladzhyan}, \citenamefont {Zyuzin},\ and\ \citenamefont
		{Simon}}]{PhysRevB.99.165302}%
	\BibitemOpen
	\bibfield  {author} {\bibinfo {author} {\bibfnamefont {V.}~\bibnamefont
			{Kaladzhyan}}, \bibinfo {author} {\bibfnamefont {A.~A.}\ \bibnamefont
			{Zyuzin}}, \ and\ \bibinfo {author} {\bibfnamefont {P.}~\bibnamefont
			{Simon}},\ }\href {\doibase 10.1103/PhysRevB.99.165302} {\bibfield  {journal}
		{\bibinfo  {journal} {Phys. Rev. B}\ }\textbf {\bibinfo {volume} {99}},\
		\bibinfo {pages} {165302} (\bibinfo {year} {2019})}\BibitemShut {NoStop}%
	\bibitem [{\citenamefont {Mastrogiuseppe}\ \emph {et~al.}(2016)\citenamefont
		{Mastrogiuseppe}, \citenamefont {Sandler},\ and\ \citenamefont
		{Ulloa}}]{PhysRevB.93.094433}%
	\BibitemOpen
	\bibfield  {author} {\bibinfo {author} {\bibfnamefont {D.}~\bibnamefont
			{Mastrogiuseppe}}, \bibinfo {author} {\bibfnamefont {N.}~\bibnamefont
			{Sandler}}, \ and\ \bibinfo {author} {\bibfnamefont {S.~E.}\ \bibnamefont
			{Ulloa}},\ }\href {\doibase 10.1103/PhysRevB.93.094433} {\bibfield  {journal}
		{\bibinfo  {journal} {Phys. Rev. B}\ }\textbf {\bibinfo {volume} {93}},\
		\bibinfo {pages} {094433} (\bibinfo {year} {2016})}\BibitemShut {NoStop}%
	\bibitem [{\citenamefont {Hosseini}\ and\ \citenamefont
		{Askari}(2015)}]{PhysRevB.92.224435}%
	\BibitemOpen
	\bibfield  {author} {\bibinfo {author} {\bibfnamefont {M.~V.}\ \bibnamefont
			{Hosseini}}\ and\ \bibinfo {author} {\bibfnamefont {M.}~\bibnamefont
			{Askari}},\ }\href {\doibase 10.1103/PhysRevB.92.224435} {\bibfield
		{journal} {\bibinfo  {journal} {Phys. Rev. B}\ }\textbf {\bibinfo {volume}
			{92}},\ \bibinfo {pages} {224435} (\bibinfo {year} {2015})}\BibitemShut
	{NoStop}%
	\bibitem [{\citenamefont {Saremi}(2007)}]{PhysRevB.76.184430}%
	\BibitemOpen
	\bibfield  {author} {\bibinfo {author} {\bibfnamefont {S.}~\bibnamefont
			{Saremi}},\ }\href {\doibase 10.1103/PhysRevB.76.184430} {\bibfield
		{journal} {\bibinfo  {journal} {Phys. Rev. B}\ }\textbf {\bibinfo {volume}
			{76}},\ \bibinfo {pages} {184430} (\bibinfo {year} {2007})}\BibitemShut
	{NoStop}%
	\bibitem [{\citenamefont {Hwang}\ and\ \citenamefont
		{Das~Sarma}(2008)}]{PhysRevLett.101.156802}%
	\BibitemOpen
	\bibfield  {author} {\bibinfo {author} {\bibfnamefont {E.~H.}\ \bibnamefont
			{Hwang}}\ and\ \bibinfo {author} {\bibfnamefont {S.}~\bibnamefont
			{Das~Sarma}},\ }\href {\doibase 10.1103/PhysRevLett.101.156802} {\bibfield
		{journal} {\bibinfo  {journal} {Phys. Rev. Lett.}\ }\textbf {\bibinfo
			{volume} {101}},\ \bibinfo {pages} {156802} (\bibinfo {year}
		{2008})}\BibitemShut {NoStop}%
	\bibitem [{\citenamefont {Black-Schaffer}(2010)}]{PhysRevB.81.205416}%
	\BibitemOpen
	\bibfield  {author} {\bibinfo {author} {\bibfnamefont {A.~M.}\ \bibnamefont
			{Black-Schaffer}},\ }\href {\doibase 10.1103/PhysRevB.81.205416} {\bibfield
		{journal} {\bibinfo  {journal} {Phys. Rev. B}\ }\textbf {\bibinfo {volume}
			{81}},\ \bibinfo {pages} {205416} (\bibinfo {year} {2010})}\BibitemShut
	{NoStop}%
	\bibitem [{\citenamefont {Sherafati}\ and\ \citenamefont
		{Satpathy}(2011)}]{PhysRevB.83.165425}%
	\BibitemOpen
	\bibfield  {author} {\bibinfo {author} {\bibfnamefont {M.}~\bibnamefont
			{Sherafati}}\ and\ \bibinfo {author} {\bibfnamefont {S.}~\bibnamefont
			{Satpathy}},\ }\href {\doibase 10.1103/PhysRevB.83.165425} {\bibfield
		{journal} {\bibinfo  {journal} {Phys. Rev. B}\ }\textbf {\bibinfo {volume}
			{83}},\ \bibinfo {pages} {165425} (\bibinfo {year} {2011})}\BibitemShut
	{NoStop}%
	\bibitem [{\citenamefont {Braunecker}\ \emph {et~al.}(2009)\citenamefont
		{Braunecker}, \citenamefont {Simon},\ and\ \citenamefont
		{Loss}}]{PhysRevLett.102.116403}%
	\BibitemOpen
	\bibfield  {author} {\bibinfo {author} {\bibfnamefont {B.}~\bibnamefont
			{Braunecker}}, \bibinfo {author} {\bibfnamefont {P.}~\bibnamefont {Simon}}, \
		and\ \bibinfo {author} {\bibfnamefont {D.}~\bibnamefont {Loss}},\ }\href
	{\doibase 10.1103/PhysRevLett.102.116403} {\bibfield  {journal} {\bibinfo
			{journal} {Phys. Rev. Lett.}\ }\textbf {\bibinfo {volume} {102}},\ \bibinfo
		{pages} {116403} (\bibinfo {year} {2009})}\BibitemShut {NoStop}%
	\bibitem [{\citenamefont {Klinovaja}\ and\ \citenamefont
		{Loss}(2013)}]{PhysRevB.87.045422}%
	\BibitemOpen
	\bibfield  {author} {\bibinfo {author} {\bibfnamefont {J.}~\bibnamefont
			{Klinovaja}}\ and\ \bibinfo {author} {\bibfnamefont {D.}~\bibnamefont
			{Loss}},\ }\href {\doibase 10.1103/PhysRevB.87.045422} {\bibfield  {journal}
		{\bibinfo  {journal} {Phys. Rev. B}\ }\textbf {\bibinfo {volume} {87}},\
		\bibinfo {pages} {045422} (\bibinfo {year} {2013})}\BibitemShut {NoStop}%
	\bibitem [{\citenamefont {Zhu}\ \emph {et~al.}(2010)\citenamefont {Zhu},
		\citenamefont {Chang}, \citenamefont {Liu},\ and\ \citenamefont
		{Lin}}]{PhysRevB.81.113302}%
	\BibitemOpen
	\bibfield  {author} {\bibinfo {author} {\bibfnamefont {J.-J.}\ \bibnamefont
			{Zhu}}, \bibinfo {author} {\bibfnamefont {K.}~\bibnamefont {Chang}}, \bibinfo
		{author} {\bibfnamefont {R.-B.}\ \bibnamefont {Liu}}, \ and\ \bibinfo
		{author} {\bibfnamefont {H.-Q.}\ \bibnamefont {Lin}},\ }\href {\doibase
		10.1103/PhysRevB.81.113302} {\bibfield  {journal} {\bibinfo  {journal} {Phys.
				Rev. B}\ }\textbf {\bibinfo {volume} {81}},\ \bibinfo {pages} {113302}
		(\bibinfo {year} {2010})}\BibitemShut {NoStop}%
	\bibitem [{\citenamefont {Schulz}\ \emph {et~al.}(2009)\citenamefont {Schulz},
		\citenamefont {De~Martino}, \citenamefont {Ingenhoven},\ and\ \citenamefont
		{Egger}}]{PhysRevB.79.205432}%
	\BibitemOpen
	\bibfield  {author} {\bibinfo {author} {\bibfnamefont {A.}~\bibnamefont
			{Schulz}}, \bibinfo {author} {\bibfnamefont {A.}~\bibnamefont {De~Martino}},
		\bibinfo {author} {\bibfnamefont {P.}~\bibnamefont {Ingenhoven}}, \ and\
		\bibinfo {author} {\bibfnamefont {R.}~\bibnamefont {Egger}},\ }\href
	{\doibase 10.1103/PhysRevB.79.205432} {\bibfield  {journal} {\bibinfo
			{journal} {Phys. Rev. B}\ }\textbf {\bibinfo {volume} {79}},\ \bibinfo
		{pages} {205432} (\bibinfo {year} {2009})}\BibitemShut {NoStop}%
	\bibitem [{\citenamefont {Schwabe}\ \emph {et~al.}(1996)\citenamefont
		{Schwabe}, \citenamefont {Elliott},\ and\ \citenamefont
		{Wingreen}}]{PhysRevB.54.12953}%
	\BibitemOpen
	\bibfield  {author} {\bibinfo {author} {\bibfnamefont {N.~F.}\ \bibnamefont
			{Schwabe}}, \bibinfo {author} {\bibfnamefont {R.~J.}\ \bibnamefont
			{Elliott}}, \ and\ \bibinfo {author} {\bibfnamefont {N.~S.}\ \bibnamefont
			{Wingreen}},\ }\href {\doibase 10.1103/PhysRevB.54.12953} {\bibfield
		{journal} {\bibinfo  {journal} {Phys. Rev. B}\ }\textbf {\bibinfo {volume}
			{54}},\ \bibinfo {pages} {12953} (\bibinfo {year} {1996})}\BibitemShut
	{NoStop}%
	\bibitem [{\citenamefont {David}\ and\ \citenamefont
		{Cserti}(2010)}]{david2010general}%
	\BibitemOpen
	\bibfield  {author} {\bibinfo {author} {\bibfnamefont {G.}~\bibnamefont
			{David}}\ and\ \bibinfo {author} {\bibfnamefont {J.}~\bibnamefont {Cserti}},\
	}\href@noop {} {\bibfield  {journal} {\bibinfo  {journal} {Physical Review
				B}\ }\textbf {\bibinfo {volume} {81}},\ \bibinfo {pages} {121417} (\bibinfo
		{year} {2010})}\BibitemShut {NoStop}%
	\bibitem [{\citenamefont {Zawadzki}\ and\ \citenamefont
		{Rusin}(2011)}]{ZB:review1}%
	\BibitemOpen
	\bibfield  {author} {\bibinfo {author} {\bibfnamefont {W.}~\bibnamefont
			{Zawadzki}}\ and\ \bibinfo {author} {\bibfnamefont {T.~M.}\ \bibnamefont
			{Rusin}},\ }\href@noop {} {\bibfield  {journal} {\bibinfo  {journal} {Journal
				of Physics: Condensed Matter}\ }\textbf {\bibinfo {volume} {23}},\ \bibinfo
		{pages} {143201} (\bibinfo {year} {2011})}\BibitemShut {NoStop}%
	\bibitem [{\citenamefont {Wang}\ \emph {et~al.}(2015)\citenamefont {Wang},
		\citenamefont {Lian},\ and\ \citenamefont {Zhang}}]{vanVleck}%
	\BibitemOpen
	\bibfield  {author} {\bibinfo {author} {\bibfnamefont {J.}~\bibnamefont
			{Wang}}, \bibinfo {author} {\bibfnamefont {B.}~\bibnamefont {Lian}}, \ and\
		\bibinfo {author} {\bibfnamefont {S.-C.}\ \bibnamefont {Zhang}},\ }\href
	{\doibase 10.1103/PhysRevLett.115.036805} {\bibfield  {journal} {\bibinfo
			{journal} {Phys. Rev. Lett.}\ }\textbf {\bibinfo {volume} {115}},\ \bibinfo
		{pages} {036805} (\bibinfo {year} {2015})}\BibitemShut {NoStop}%
	\bibitem [{\citenamefont {Liu}\ \emph {et~al.}(2019)\citenamefont {Liu},
		\citenamefont {Xu}, \citenamefont {He}, \citenamefont {van~der Laan},
		\citenamefont {Zhang},\ and\ \citenamefont {Wang}}]{vanVleck2}%
	\BibitemOpen
	\bibfield  {author} {\bibinfo {author} {\bibfnamefont {W.}~\bibnamefont
			{Liu}}, \bibinfo {author} {\bibfnamefont {Y.}~\bibnamefont {Xu}}, \bibinfo
		{author} {\bibfnamefont {L.}~\bibnamefont {He}}, \bibinfo {author}
		{\bibfnamefont {G.}~\bibnamefont {van~der Laan}}, \bibinfo {author}
		{\bibfnamefont {R.}~\bibnamefont {Zhang}}, \ and\ \bibinfo {author}
		{\bibfnamefont {K.}~\bibnamefont {Wang}},\ }\href@noop {} {\bibfield
		{journal} {\bibinfo  {journal} {Science advances}\ }\textbf {\bibinfo
			{volume} {5}},\ \bibinfo {pages} {eaav2088} (\bibinfo {year}
		{2019})}\BibitemShut {NoStop}%
	\bibitem [{\citenamefont {Li}\ \emph {et~al.}(2015)\citenamefont {Li},
		\citenamefont {Chang}, \citenamefont {Wu}, \citenamefont {Tao}, \citenamefont
		{Zhao}, \citenamefont {Chan}, \citenamefont {Moodera}, \citenamefont {Li},\
		and\ \citenamefont {Zhu}}]{vanVleck3}%
	\BibitemOpen
	\bibfield  {author} {\bibinfo {author} {\bibfnamefont {M.}~\bibnamefont
			{Li}}, \bibinfo {author} {\bibfnamefont {C.-Z.}\ \bibnamefont {Chang}},
		\bibinfo {author} {\bibfnamefont {L.}~\bibnamefont {Wu}}, \bibinfo {author}
		{\bibfnamefont {J.}~\bibnamefont {Tao}}, \bibinfo {author} {\bibfnamefont
			{W.}~\bibnamefont {Zhao}}, \bibinfo {author} {\bibfnamefont {M.~H.~W.}\
			\bibnamefont {Chan}}, \bibinfo {author} {\bibfnamefont {J.~S.}\ \bibnamefont
			{Moodera}}, \bibinfo {author} {\bibfnamefont {J.}~\bibnamefont {Li}}, \ and\
		\bibinfo {author} {\bibfnamefont {Y.}~\bibnamefont {Zhu}},\ }\href {\doibase
		10.1103/PhysRevLett.114.146802} {\bibfield  {journal} {\bibinfo  {journal}
			{Phys. Rev. Lett.}\ }\textbf {\bibinfo {volume} {114}},\ \bibinfo {pages}
		{146802} (\bibinfo {year} {2015})}\BibitemShut {NoStop}%
\end{thebibliography}
%

\onecolumngrid

\appendix

\section{Involution operators}\label{App1}

The decomposition of the Hamiltonian is represented by following involution operators
\begin{eqnarray}
{\hat T}&&=\left(
\begin{array}{cccc}
-\sin (\Theta ) & 0 & 0 & -\cos (\Theta ) \\
0 & -\sin (\Theta ) & -\cos (\Theta ) & 0 \\
0 & -\cos (\Theta ) & \sin (\Theta ) & 0 \\
-\cos (\Theta ) & 0 & 0 & \sin (\Theta ) \\
\end{array}
\right),
\end{eqnarray}
and
\begin{equation}
{\hat R}=
\left(\begin{array}{cccc}
0&A&C&0\\
A^*&0&0&C^*\\
C^*&0&0&B^*\\
0&C&B&0
\end{array}\right)
\end{equation}
where $A=bk_u\cos\Theta^2+(a+b\sin\Theta)(ik_ve^{i\Phi}+k_u\sin\Theta), B=bk_u\cos\Theta^2+(a-b\sin\Theta)(ik_ve^{i\Phi}-k_u\sin\Theta), C=(ak_u+ibe^{i\Phi}k_v)\cos\Theta$ in which $a=-\frac{E_+-E_-}{2E_+E_-}, b=-\frac{E_++E_-}{2E_+E_-}$, and $E_\pm=E_{+\pm}$.
These expressions can be derived by noting from Eq. \eqref{Qst} that the projectors $Q_{s,\tau}$s are linear sums of identity operators, $\hat{R}$, $\hat{T}$, and $\hat{R}\hat{T}$. The appropriate linear combination of the four $\hat{Q}_{s,\tau}$s that would yield $\hat{R}$ and $\hat{T}$ can then be easily solved for to obtain $\hat{T} = \hat{Q}_{++} -\hat{Q}_{+-}+\hat{Q}_{-+}-\hat{Q}_{--}$ and $\hat{R} = \hat{Q}_{++} +\hat{Q}_{+-}-\hat{Q}_{-+}-\hat{Q}_{--}$.

\section{Green's function}\label{App2}
We consider two magnetic centers $S_i$ ($i=1,2$) located at $\bm R_i$. For simplicity, we assume that $\bm {R}_1=\bm 0$, and $\bm {R}_2= \bm R$.
In the framework of second order perturbation theory, the effective interaction between two magnetic impurities is given by \cite{RKKY1,RKKY2,RKKY3,mattis2006theory}
\begin{eqnarray}
H_{RKKY}=-\frac{J^2}{\pi}\times \mathrm{Tr}\left[\int_{-\infty}^{\epsilon_F}d\epsilon \mathrm{Im}\left\{({\bm S}_1\cdot{\tilde {\bm \sigma}})G({\bm R},\epsilon^+)({\bm S}_2\cdot{\tilde {\bm \sigma}})G(-{\bm R},\sigma^+)\right\}\right],
\end{eqnarray}
where $\epsilon=E+i0^+$, the spin operator is defined as ${\tilde {\bm \sigma}}=\tau_0\otimes{\bm \sigma}$, and Tr stands for tracing over the spin degree of freedom. The Green's function in real space is given by the Fourier transformation
\begin{eqnarray}
G({\bm R},\epsilon^+)=\int \frac{d^2 q}{A_{BZ}}e^{i{\bm q}\cdot {\bm R}}G({\bm q},\epsilon^+),
\end{eqnarray}
where $G({\bm q},\epsilon^+)=[\epsilon^+-H_0({\bm q})]^{-1}$ is the Green's function in momentum space, and $A_{BZ}$ is the area of the first Brillouin zone. Explicitly, the Green's function be be written in block matrix form as
\begin{eqnarray}
G({\bm q},\epsilon)=\begin{bmatrix}
    G_{+} & G_{12}\\
    G_{21} & G_{-}
\end{bmatrix}.
\end{eqnarray}

\subsection{Green's function in the limit $\Delta\ll\epsilon_B$}
In the limit $\Delta\ll\epsilon_B=v_fq_0$, the Green's function matrix can be approximated as
\begin{eqnarray}
G_{+}({\bm q})=\left[\epsilon-H_D({\bm q}-{\bm q}_0)\right]^{-1},\ \  G_{-}({\bm q})=\left[\epsilon+H_D({\bm q}+{\bm q}_0)\right]^{-1}\\
G_{12}({\bm q})=\Delta G_{+}({\bm q})G_{-}({\bm q}),\ \ G_{21}(q)=\Delta G_{-}({\bm q})G_{+}({\bm q}).
\end{eqnarray}
Explicitly,
\begin{eqnarray}
G_{\pm}({\bm q})=\frac{1 }{\epsilon ^2-v_f^2 q_\pm^2}\left[\epsilon\mp H_D({\bm q}_\pm)\right],
\end{eqnarray}
where ${\bm q}_\pm={\bm q}\mp {\bm q}_0$.

The real space Green's function can be obtained: 
\begin{eqnarray}
G_\pm({\bm R},\epsilon^+)=-\frac{2\pi e^{\pm i{\bm R}\cdot{\bm q}_0}}{v_f^2A_{BZ}}\left[A\pm B({\hat{\bm R}}\times{\bm \sigma})_z\right],\\
\end{eqnarray}
where $A=\epsilon^+ K_0(-i|\eta|)$ and  $B=|\epsilon^+|K_1(-i|\eta|)$ with $\eta=\frac{R\epsilon^+}{v_f}$, $K_n(\eta)$ is the modified Bessel function of the second kind, and ${\hat{\bm R}}={{\bm R}}/R$ is the unit vector along the ${\bm R}$. The off-diagonal Green's functions can be expressed as the convolutions
\begin{eqnarray}
G_{12}({\bm R})=\Delta \int{\frac{d{\bm R}'}{V} G_+({\bm R}')G_-({\bm R}-{\bm R}')},\\
G_{21}({\bm R})=\Delta \int{\frac{d{\bm R}'}{V} G_-({\bm R}-{\bm R}')G_+({\bm R}')},\\
\end{eqnarray}
which satisfy $G_{12}(R)^\dagger=G_{21}(-R)$. Substituting the expression of $G_\pm$ into the above, we obtain
\begin{eqnarray}
G_{12}({\bm R})=\Delta e^{- i{\bm R}\cdot{\bm q}_0}
\left(\frac{2\pi}{v_f^2A_{BZ}}\right)^2\int\frac{d{\bm R}'}{V} e^{i2{\bm R}'\cdot{\bm q}_0}\left[A_{r}+B_{r}({\hat{\bm r}}\times{\bm \sigma})_z\right]\left[A_{R-r}+B_{R-r}({\hat{\bm R-r}}\times{\bm \sigma})_z\right],\\
G_{21}({\bm R})=\Delta e^{- i{\bm R}\cdot{\bm q}_0}
\left(\frac{2\pi}{v_f^2A_{BZ}}\right)^2\int\frac{d{\bm R}'}{V} e^{i2{\bm R}'\cdot{\bm q}_0}\left[A_{R-r}+B_{R-r}({\hat{\bm R-r}}\times{\bm \sigma})_z\right]\left[A_{r}+B_{r}({\hat{\bm r}}\times{\bm \sigma})_z\right].
\end{eqnarray}
Disregarding the off-diagonal elements $G_{12},G_{21}$, we have
\begin{eqnarray}\label{RKKY}
H_{RKKY}=F_1({\bm S}_1\cdot {\bm S}_2)+F_2({\bm S}_1\cdot {\hat{\bm u}})({\bm S}_2 \cdot {\hat{\bm u}})+F_3\left[{\hat{\bm u}}\cdot({\bm S}_1\times {\bm S}_2)\right],
\end{eqnarray}
in which ${\hat{\bm u}}=({\hat{\bm R}}\times {\bm z})$, and the range functions are
\begin{eqnarray}
F_1&&=I_1\cos(2{\bm R}\cdot {\bm q}_0),\\
F_2&&=I_2\cos(2{\bm R}\cdot {\bm q}_0),\\
F_3&&=I_3\sin(2{\bm R}\cdot {\bm q}_0),
\end{eqnarray}
where
\begin{eqnarray}
I_1&&=-\frac{16\pi J^2}{v_f^4A_{BZ}^2} \mathrm{Im}\int_{-\infty}^{\epsilon_f}d\epsilon (A^2+B^2) e^{\lambda \epsilon},\\
I_2&&=\frac{32\pi J^2}{v_f^4A_{BZ}^2} \mathrm{Im}\int_{-\infty}^{\epsilon_f}d\epsilon (B^2)e^{\lambda \epsilon},\\
I_3&&=-\frac{32\pi J^2}{v_f^4A_{BZ}^2} \mathrm{Im}\int_{-\infty}^{\epsilon_f}d\epsilon (AB)e^{\lambda \epsilon}.
\end{eqnarray}
In the above, $e^{\lambda \epsilon}$ is a smooth cutoff function with $\lambda\rightarrow 0^+$ to avoid the divergence in integrating.
Changing the integration variable to $\epsilon=v_f\eta/R$, the above integrals can be rewritten as $I_i\propto \frac{4\pi J^2}{v_fA_{BZ}^2R^3} \int_{-\infty}^{\eta_f}d\eta f(\eta)$, where $\eta=R\epsilon_f/v_f$. When $\epsilon_f=0$ and $\eta_f=0$, the integral $\int_{-\infty}^{\eta_f}d\eta f(\eta)$ converges to a constant number so that the range functions depend on the distance as $R^{-3}$. For $\eta\gg 1$, we have approximations $\mathrm{Im}(A^2)=\mathrm{Im}(B^2)=\mathrm{Im}(AB)\approx \pm \frac{\pi\cos(2\eta)}{2|\eta|}$. With the above, we can obtain the asymptotic range functions as
\begin{eqnarray}
I_1=I_3=-I_2=&&-\frac{32\pi J^2}{v_fA_{BZ}^2R^3} \frac{\pi}{4}\left[\cos\frac{2R\epsilon_f}{v_f}+\frac{2R\epsilon_f}{v_f}\sin\frac{2R\epsilon_f}{v_f}-2\right]\nonumber\\
&&\approx-\frac{64\pi J^2\epsilon_f}{v_f^2A_{BZ}^2R^2} \sin\frac{2R\epsilon_f}{v_f}
\end{eqnarray}
which decay as $R^{-2}$. We note here that when only one surface state contributes to the RKKY coupling, one can set ${\bm q}_0=0$ so that the spin-frustrated term vanishes, which is consistent with previous results \cite{RKKYTI:PRL2011}.

The RKKY coupling consists of three terms: the Heisenberg, the Dzyaloshinsky-Moriya, and spin-frustrated terms. As can be seen, the RKKY couplings decay with distance as $R^{-2}$, which is consistent with previous results \cite{RKKYTI:PRL2011}. Furthermore, they are also anisotropic with respect to the direction between the two magnetic impurities, as can be seen from the $\cos(2{\bm R}\cdot {\bm q}_0)$ terms in the range functions.

\subsection{Green's function in the limit $\Delta\gg\epsilon_B$}

Int the limit $\Delta\gg v_f q_0$, there is a gap opening of $2\Delta$ at the $\Gamma-$point. In this case, the Green's functions can be obtained as
\begin{eqnarray}
G({\bm q})&&=-\frac{1 }{\Delta ^2-\epsilon ^2+q^2 v_f^2}\left[\epsilon+H_0({\bm q})\right].
\end{eqnarray}
The real space Green's function obtained for $R\gg 1$ is
\begin{eqnarray}
G({\bm R})&&=-\int_0^\infty \frac{dq}{A_{BZ}}\frac{2\pi q}{w^2+q^2v_f^2}\left\{J_0(qR)(\epsilon+\tau_x\Delta)+i \tau_zJ_1(qR)v_fq({\bm z}\times {\hat{\bm R}})\cdot{\bm\sigma}\right\},\\
&&=- \frac{2\pi}{A_{BZ}}\frac{1}{v_f^2}\left\{K_0(\frac{R|w|}{v_f})(\epsilon+\tau_x\Delta)+i \tau_z|w|K_1(\frac{R|w|}{v_f})({\bm z}\times {\hat{\bm R}})\cdot{\bm\sigma}\right\},
\end{eqnarray}
in which $w=\sqrt{\Delta^2-\epsilon^2}$, and $J_n(x)$ is the Bessel function of the first kind. 

The RKKY coupling reads as
\begin{eqnarray}
H_{RKKY}={\tilde F}_1({\bm S}_1\cdot {\bm S}_2)+{\tilde F}_2({\bm S}_1\cdot {\hat{\bm u}})({\bm S}_2 \cdot {\hat{\bm u}})+{\tilde F}_3\left[{\hat{\bm u}}\cdot({\bm S}_1\times {\bm S}_2)\right],
\end{eqnarray}
where the range functions are now given by
\begin{eqnarray}
{\tilde F}_1&&=-\frac{16\pi J^2}{v_f^4A_{BZ}^2} \mathrm{Im}\int_{-\infty}^{\epsilon_f} ({\tilde A}_1^2+{\tilde A}_2^2-{\tilde B}^2),\\
{\tilde F}_2&&=-\frac{32\pi J^2}{v_f^4A_{BZ}^2}\mathrm{Im}\int_{-\infty}^{\epsilon_f} {\tilde B}^2,\\
{\tilde F}_3&&=\frac{32\pi J^2}{v_f^4A_{BZ}^2}\mathrm{Im}\int_{-\infty}^{\epsilon_f}  {\tilde A}_1{\tilde B},\\
\end{eqnarray}
where ${\tilde A}_1=\epsilon K_0(\frac{R|w|}{v_f}),
{\tilde A}_2=\Delta K_0(\frac{R|w|}{v_f}),
{\tilde B}=|w|K_1(\frac{R|w|}{v_f}).$

\end{document}